\documentclass[11pt,a4paper]{article}
\pdfoutput=1 
\usepackage{url}
\usepackage{jcappub}
\usepackage{amsfonts}
\usepackage{subfigure}
\usepackage{graphicx}

\newcommand{\be}{\begin{equation}}
\newcommand{\ee}{\end{equation}}
\newcommand{\bea}{\begin{eqnarray}}
\newcommand{\eea}{\end{eqnarray}}

\newcommand{\beq}{\begin{equation}}
\newcommand{\eeq}{\end{equation}}

\begin{document}

%%%%%%%%%%%%%%%%%%%%%%%%%%%%%%%%%%%%%%%%%%%%%%%%%%%%%%%%%%%%%%%%%%%%%%
% Frontpage %%%%%%%%%%%%%%%%%%%%%%%%%%%%%%%%%%%%%%%%%%%%%%%%%%%%%%%%%%
%%%%%%%%%%%%%%%%%%%%%%%%%%%%%%%%%%%%%%%%%%%%%%%%%%%%%%%%%%%%%%%%%%%%%%

%\subheader{\hfill}

\title{Heavy sterile neutrino emission in core-collapse supernovae: Constraints and signatures}

\author[a]{Leonardo Mastrototaro,}
\author[a,b]{Alessandro Mirizzi,}
\author[c]{Pasquale Dario Serpico,}
\author[d]{Arman Esmaili}

\affiliation[a]{Dipartimento Interateneo di Fisica ``Michelangelo Merlin'', Via Amendola 173, 70126 Bari, Italy}
\affiliation[b]{Istituto Nazionale di Fisica Nucleare - Sezione di Bari, Via Amendola 173, 70126 Bari, Italy}
\affiliation[c]{LAPTh,  Univ.   Grenoble  Alpes,  USMB,  CNRS,  F-74000  Annecy,  France}
\affiliation[d]{Departamento de F\'isica, Pontif\'icia Universidade Cat\'olica do Rio de Janeiro, Rio de Janeiro 22452-970, Brazil}
\emailAdd{l.mastrototaro5@studenti.uniba.it, alessandro.mirizzi@ba.infn.it, serpico@lapth.cnrs.fr, arman@puc-rio.br }

\abstract{Heavy sterile neutrinos with masses ${\mathcal O}(100)$~MeV mixing with active neutrinos can be produced in the core
of a collapsing supernova (SN). In order to avoid an excessive energy loss, shortening the observed duration of the SN 1987A neutrino burst, we show that the active-sterile neutrino mixing
angle should satisfy $\sin^2 \theta \lesssim 5 \times 10^{-7}$. For a mixing with tau flavour, this bound
is much stronger than the ones from laboratory searches. Moreover, we show that  in the viable parameter space the decay of  such  ``heavy'' sterile neutrinos in the SN envelope would lead to 
a very energetic flux of daughter active neutrinos; if not too far below current limits, this would be detectable in  large underground neutrino observatories, like Super-Kamiokande, as a (slightly time-delayed) high-energy bump in the spectrum of a forthcoming Galactic SN event.}
\hfill{\small LAPTH-047/19}\\
\maketitle

%%%%%%%%%%%%%%%%%%%%%%%%%%%%%%%%%%%%%%%%%%%%%%%%%%%%%%%%%%%%%%%%%%%%%%%%%%%%%%%%%%%%%%%%%
\section{Introduction}
%....................................................................
Core-collapse supernovae (SNe) represent the most powerful sources of neutrinos in the Universe. Indeed, during the explosion of a massive star,
${\mathcal O} (10^{58})$ (anti)neutrinos of  all the flavors  are emitted with average energies $E \sim 15$~MeV. For this unique property, a SN 
can be truly considered a  \emph{neutrino factory} to probe fundamental properties of neutrinos, e.g.  mixing and nonstandard
interactions  (see, e.g.,~\cite{Mirizzi:2015eza,Horiuchi:2017sku} for recent reviews). At this regard,  the next detection of a SN neutrino burst from a Galactic event in a large underground detector
is considered  one of the next frontiers of low-energy neutrino astronomy. Furthermore, a SN is also a powerful laboratory to probe the emission of exotic particles beyond standard Model neutrinos. In particular, low-mass
particles produced in the SN core would constitute a novel channel of \emph{energy loss}, shortening the duration of
 the neutrino burst~\cite{Raffelt:2012kt}. 
In this context, the detection of SN 1987A neutrinos on a time scale of  $\sim 10$~s, as expected from the Standard
SN cooling scenario, excludes an additional  energy drain
associated with exotic particles, if dominant with respect to the standard neutrino channels.
Many new physics scenarios have been constrained using this powerful argument, 
including axions~\cite{Carenza:2019pxu}, dark photons~\cite{Chang:2016ntp}, Kaluza-Klein gravitons~\cite{Hannestad:2001jv} and unparticles~\cite{Hannestad:2007ys}. 

Even within the neutrino sector, the duration of the SN neutrino burst has been used to constrain \emph{sterile} states
mixing with the active ones. In particular, strong constraints have been placed on  sterile neutrinos with masses of
${\mathcal O}$(keV)~\cite{Raffelt:2011nc,Arguelles:2016uwb,Suliga:2019bsq,Syvolap:2019dat}, which may be also invoked as warm dark matter candidates.
The impact of sterile states at the eV scale, as possibly suggested by a number of laboratory anomalies, on the SN signal phenomenology has also been considered~\cite{Choubey:2006aq,Choubey:2007ga,Tamborra:2011is,Esmaili:2014gya}. 
A less well-studied case (for some dated exception, see~\cite{Dolgov:2000pj,Dolgov:2000jw}) is constituted by sterile neutrinos with masses of
${\mathcal O}(100)$~MeV. Given a typical temperature $T \simeq 30$~MeV in the SN core, such relatively heavy sterile neutrinos might 
still be produced by the mixing with the active ones, suffering only a moderate suppression due to the Boltzmann factor.  

Nonetheless, sterile neutrinos in this mass range emerge rather naturally in extensions of the Standard Model, like dynamical electroweak symmetry
breaking~\cite{Appelquist:2002me} or the Neutrino Minimal Standard Model ($\nu$MSM)~\cite{Asaka:2005an,Asaka:2005pn}. In the latter case, they are related to fundamental problems of particle physics like
the origin of neutrino mass, the baryon asymmetry in the Early Universe and the nature of dark matter.
 Their parameter space is strongly constrained by collider and beam-dump experiments
for the mixing with $\nu_e$ and $\nu_\mu$ neutrinos~\cite{Alekhin:2015byh,Chun:2019nwi}. However, for  sterile neutrino mixing with $\nu_\tau$, the range $10^{-7} \lesssim \sin^2 \theta \lesssim 10^{-5}$
is currently unconstrained for masses of  ${\mathcal O}(100-200)$~MeV. In~\cite{Fuller:2009zz} it was proposed that for viable mixings 
in this mass range
there should be a significant sterile neutrino emission from a SN core, whose further decay in the stellar envelope may be helping the explosion mechanism (a scenario recently explored in~\cite{Rembiasz:2018lok})
and produce a detectable flux of energetic active neutrinos as decay byproducts.
Motivated by this intriguing insight, we decided to take a fresh look to this physics case to assess in a more quantitative way observable signatures 
of heavy sterile neutrinos emitted during a galactic SN explosion. 
The plan of our work is as follow. In Sec.~\ref{model} we present our benchmark  model for the active SN neutrino fluxes based on state-of-the-art simulations and
we characterize the emissivity of heavy sterile neutrinos solving the transport Boltzmann equation for the mixed active-sterile neutrino system in the SN core.
We also present our bound on the heavy sterile neutrino parameter space based on the energy-loss argument from SN 1987A. 
Then in Sec.~\ref{decayflux} we characterize the flux of high-energy active neutrinos, coming from the decay of the heavy sterile ones (more technical details on the spectra of different channels are reported in Appendix~\ref{nudecayspectra}). 
Its detectability in water Cherenkov detectors such as Super-Kamiokande is investigated in Sec.~\ref{detection}. 
Finally, in Sec.~\ref{conclusions} we discuss our results and we conclude.

%%%%%%%%%%%%%%%%%%%%%%%%%
\section{Heavy sterile neutrino emission from supernovae}\label{model}
%%%%%%%%%%%%%%%%%%%%%%%%%%

%.....................................
\subsection{SN reference model}
%.........................................

%%%%%%%%%%%%%
\begin{figure}[t!]
	\vspace{0.cm}
	\hspace{1.5cm}
	\includegraphics[width=0.8\textwidth]{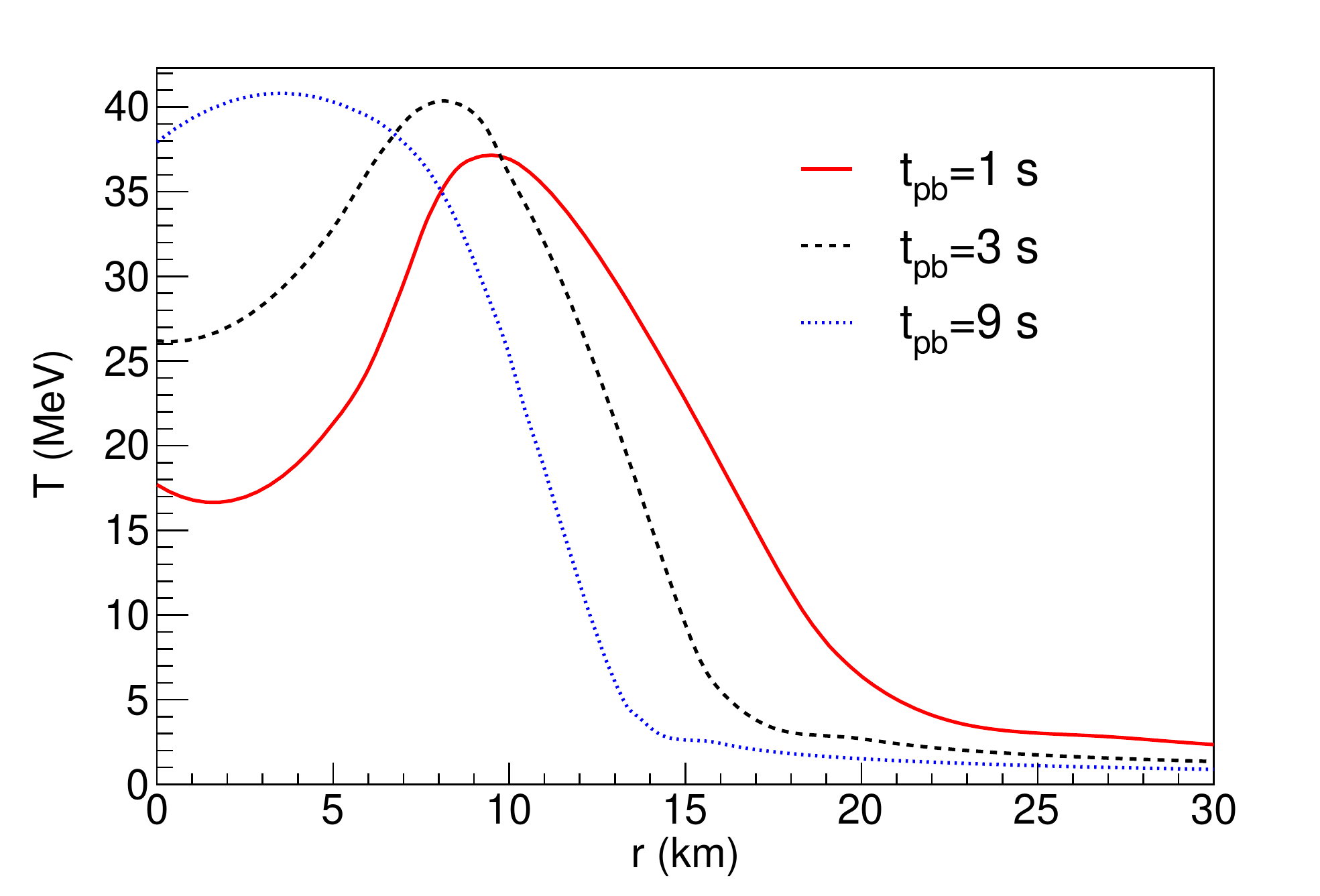}
	\caption{
Radial profile of temperature at three different post-bounce times for our  $18$~$M_{\odot}$ SN reference model.}
	\label{fig:Temp}
\end{figure}
%%%%%%%%%%%%%%%%%%%%%%%

%%%%%%%%%%%%%%%%%%%%%
\begin{figure}[!h]
\centering
\includegraphics[width=0.6\textwidth]{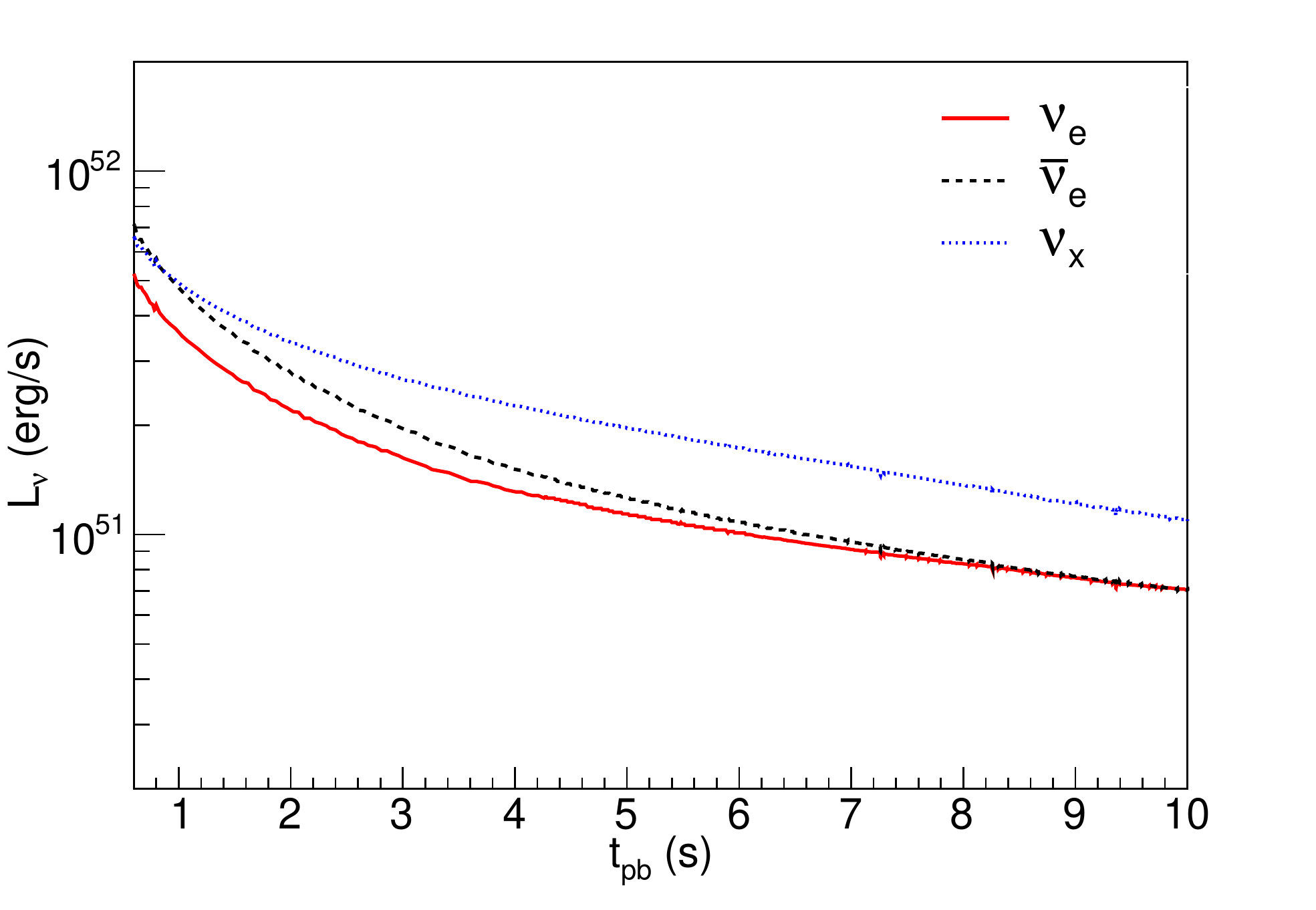}
\includegraphics[width=0.6\textwidth]{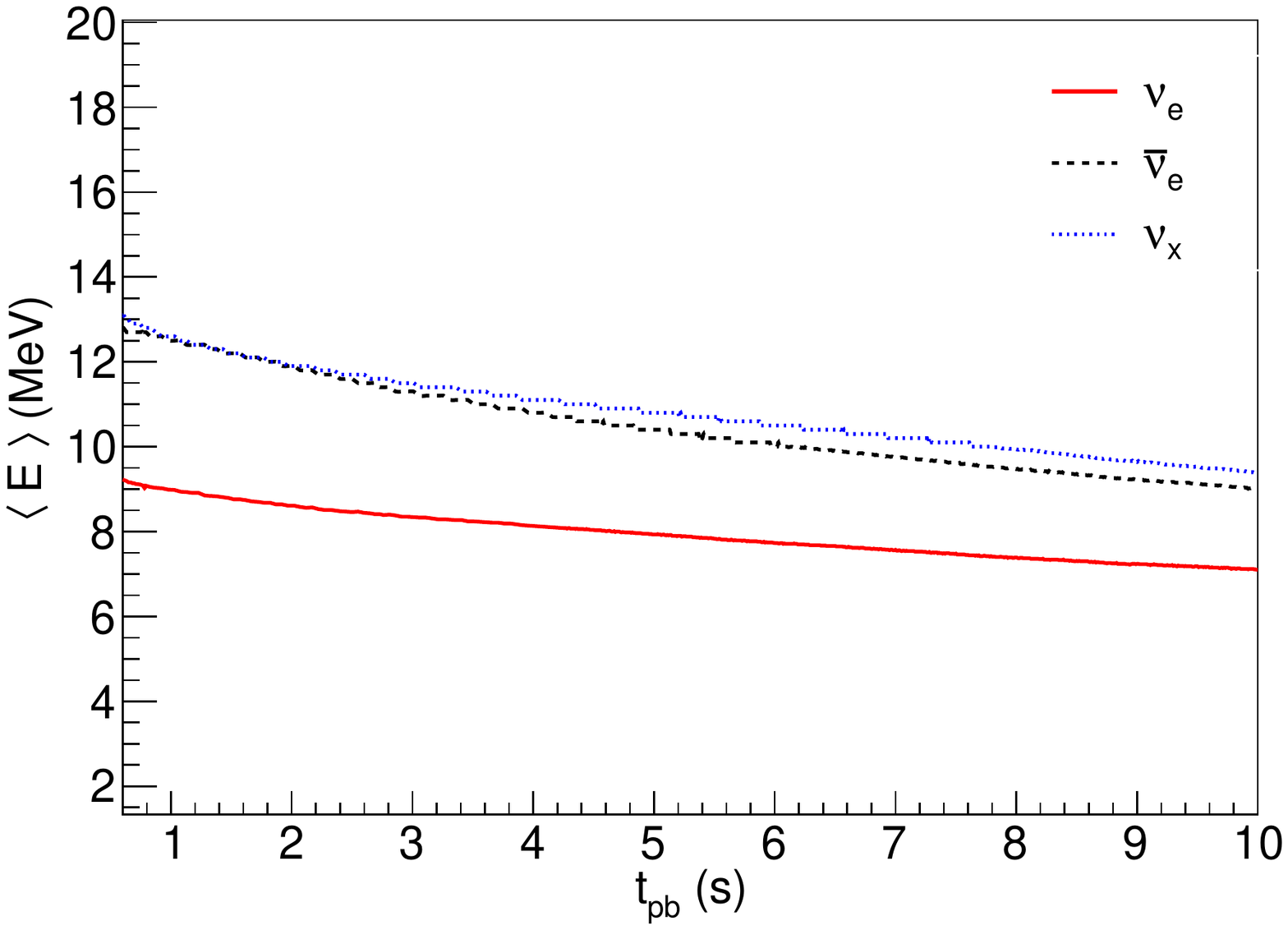}
\includegraphics[width=0.6\textwidth]{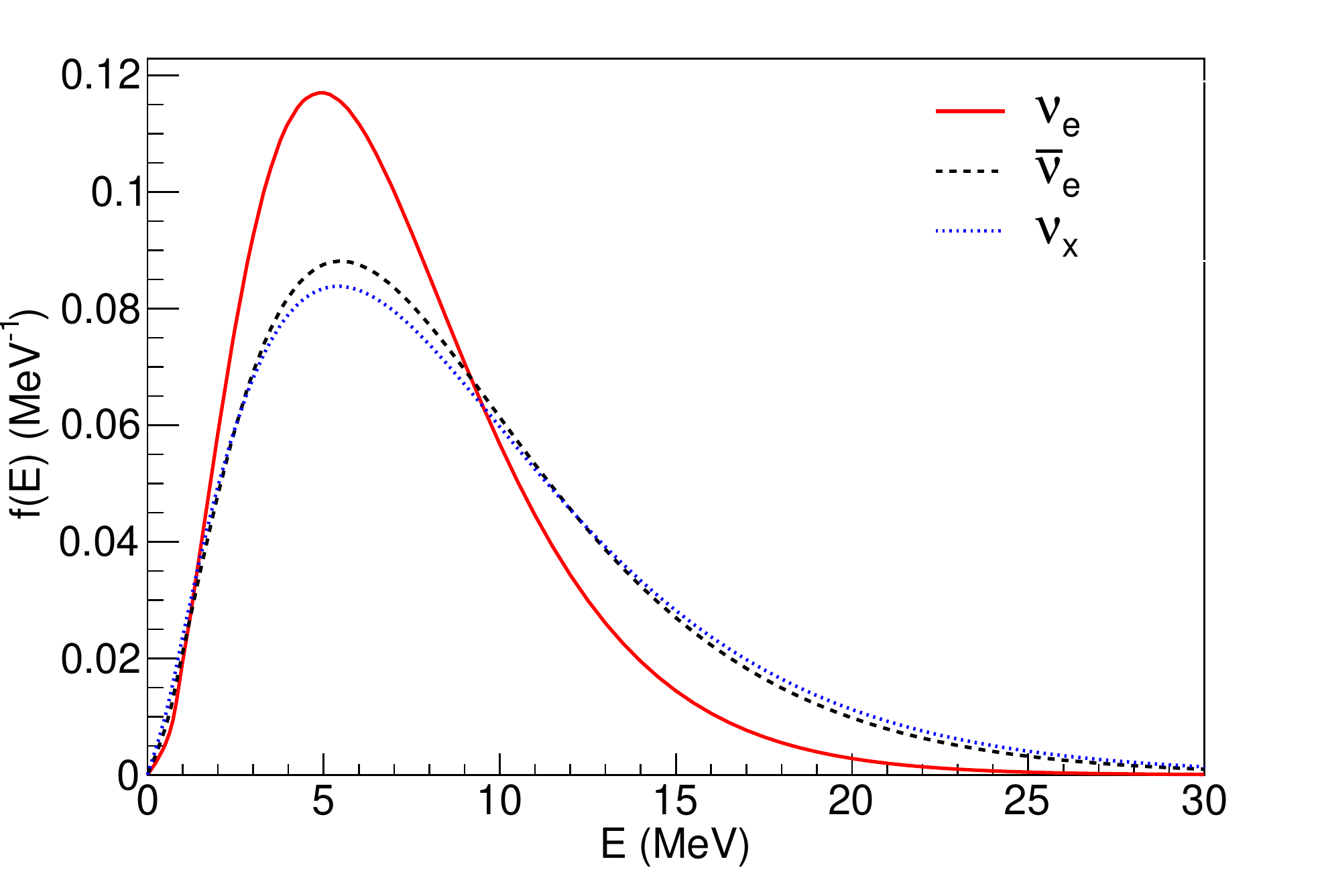}
\caption{{\it Upper panel}: Luminosities of different active 
neutrino flavours vs. the post-bounce time $t_{\rm pb}$. {\it Middle panel}:  
Average energy of different neutrinos species vs. the post-bounce time $t_{\rm pb}$.
{\it Lower panel:}  Normalized energy distribution of different neutrinos species in the post-bounce time window $[0.6,10]~\mathrm{s}$.
All panels refer to our $18~M_{\odot}$ SN progenitor benchmark. }
\label{activenu}
\end{figure}
%%%%%%%%%%%%%%%%%%%%%%%%%%%%%%%%

In this Section we characterize the heavy sterile neutrino emissivity in core-collapse supernovae, ignoring feedback onto the dynamics dominated by the standard processes.
This hypothesis will be correct for sufficiently small mixing angles, of course.  
In order to connect  these results with realistic SN models, 
we consider as SN reference a model with 18 $M_{\odot}$ progenitor 
simulated  in spherical symmetry with the  AGILE-BOLTZTRAN  code~\cite{Mezzacappa:1993gn,Liebendoerfer:2002xn}.
We remind the reader that  in spherically symmetric simulations neutrino-driven explosions
 cannot be obtained, except for very light progenitor stars. Therefore, an 
enhanced (active) neutrino heating treatment has been applied here in order to trigger the SN explosion onset at post-bounce time $t_{\rm pb} =$ 350 ms. Once the explosion proceeds,  the standard rates are restored. 
The radial temperature profile of our SN model, at three different post-bounce times $t_{\rm pb}$, shown in Fig.~\ref{fig:Temp}, corresponds to the simulation of the protoneutron star deleptonization of~\cite{Fischer:2016}.
We can see that the temperature $T$ presents a peak of $\sim 40$~MeV 
at $r \simeq 10$~km. 
Given the steep temperature dependence of the sterile neutrino emission rate (we find a  dependence on the temperature
scaling as $T^{12}$), one expects the dominant $\nu_s$ production around this peak temperature. One also realizes that at later
$t_{\rm pb}$ the peak in the temperature recedes, due to the cooling of the proto-neutron star.

Concerning the active neutrino signal, 
examining the numerical models in detail
one finds that roughly half the energy is emitted during the accretion phase ($t_{\rm pb} \lesssim 0.5$~s). 
However, it is well known that exotic processes can hardly compete in this stage~\cite{Raffelt:1987yt}. 
Conversely, the later $\nu$ signal (corresponding to the \emph{cooling} of the proto-neutron star) can be significantly
affected by the emission of new particles~\cite{Raffelt:1987yt}.
For this reason, we will focus on  $t_{\rm pb} \gtrsim 0.5$~s.
During the cooling phase, 
a SN can be roughly considered as a black-body cooling via neutrino emission. Indeed, the neutrino energy distribution is well approximated with a quasi-thermal distribution with the temperature of the neutrino-sphere. A simple parametrization of the neutrino energy distributions for the different neutrino species
$\nu \equiv \nu_e, \bar\nu_e, \nu_x, \bar\nu_x$ (with $x= \mu, \tau$), based on the numerical simulations is~\cite{Raffelt:2003en,Tamborra:2012ac}:
%%%%%%%%%%%%%%%%%%%%
\begin{align}
f_\nu (E)&=\frac{(1+\alpha_\nu)^{1+\alpha_\nu}}{\Gamma(1+\alpha_\nu)\left\langle E\right\rangle}\left(\frac{E}{\left\langle E_\nu\right\rangle}\right)^{\alpha}\exp\left[(1+\alpha_\nu)\frac{E}{\left\langle E_\nu\right\rangle}\right] \,\ , 
\label{fesuppp} \\
\frac{{\rm d}N_\nu}{{\rm d}E}&=\frac{L_\nu}{\left\langle E_\nu\right\rangle}f(E) \,\ , 
\label{miservecitare}\\
\alpha_\nu&=\frac{2\left\langle E_\nu\right\rangle ^2-\left\langle E_\nu^2\right\rangle}{\left\langle E_\nu^2\right\rangle-\left\langle E_\nu \right\rangle ^2} \,\ ,
\end{align}
%%%%%%%%%%%%%%%%%%%%%%%%%%%
where $f_\nu(E)$ is te normalized neutrino energy distribution, ${\rm d}N_\nu/{\rm d}E$ the differential number flux, $L_\nu$ is the  luminosity, and $\alpha_\nu$ is the so-called pinching parameter.

In Fig.~\ref{activenu} we plot for the different neutrino species: the neutrino luminosity $L_\nu$ vs post-bounce time $t_{\rm pb}$ (upper panel), the average neutrino energy $\left\langle E\right\rangle$ as function of the post-bounce time $t_{\rm pb}$ (middle panel) and the normalized neutrino energy distribution $f_\nu (E)$  in the time window $[0.6,10]~\mathrm{s}$  (lower panel). In Table~\ref{Bestfitparameters} we report the parameters for $\alpha_\nu$, $N_\nu \equiv L_\nu/ \langle E_\nu\rangle$, $\langle E_\nu\rangle$ referring to the considered simulation in the time window $[0.6,10]~\mathrm{s}$.\\

%%%%%%%%%%%%%%%%%%%%%%%%
\begin{table}
\caption{Parameter of the time-integrated neutrino fluxes in the time window $0.6-10~\mathrm{s}$.}
\vspace{0.5 cm}
\centering
\begin{tabular}{lccc}
\hline
%\toprule
Specie& $\alpha_\nu$& $N_\nu~ (\times 10^{56})$ & $\langle E_\nu\rangle~(\mathrm{MeV})$\\
\hline
\hline
%\midrule
$\nu_e$& $2.2$ & $9.2$ & $7.1$\\
$\bar{\nu}_e$& $1.6$ & $8.5$ & $8.8$\\
$\nu_{x}$& $1.4$ & $10.7$ & $9.1$\\
\hline
%\bottomrule
\end{tabular}
\label{Bestfitparameters}
\end{table}
%%%%%%%%%%%%%%%%%%%%%%%%%%%

%%%%%%%%%%%%%%%%%%%%%%%%%%%%
\subsection{Heavy sterile neutrinos production processes}
%%%%%%%%%%%%%%%%%%%%%%%%%%%
\label{sterileneutrinoprodution}
After having characterized the standard active neutrino emission from our benchmark core-collapse SN, we investigate the possibility that also heavy sterile neutrinos would be produced during a stellar collapse. We consider a single sterile neutrino mass eigenstate $\nu_4$, with $m_4\sim 200~\mathrm{MeV}$, and mixing angle $\theta_{a4}$. We will see that, in the chosen mass range, in order to have a sizable signal one requires $\sin^2\theta_{a4}\sim 10^{-8}-10^{-7}$. However, in the considered mass range, constraints coming from laboratory experiments and big bang nucleosynthesis would exclude these values of the mixing parameter { for the mixing with electron and muon neutrinos}~\cite{Alekhin:2015byh}. Therefore, in the following we will focus on the sole mixing of $\nu_4$ with $\nu_{\tau}$, which is the least constrained sector and where SN neutrinos have the maximal complementarity to other experiments.

First, let us consider the production processes of heavy sterile neutrinos in a SN. In the hot core, $n,p,e^-,e^+,\nu_e, \bar{\nu}_e$ are degenerate. Therefore, Pauli-blocking will render pair annihilation and the inelastic scattering of non-degenerate neutrino species $\nu_{x=\mu,\tau}$  the dominant process for the sterile $\nu_4$ mass eigenstate production. In Table~\ref{Matrixelementsterile}, we report the squared interaction matrix element (times the statistical factors $S$) for the different production processes~\cite{Hannestad:1995rs,Fuller:2009zz} in terms of Mandelstam variables. Here and in the following we assume sterile neutrinos as Majorana particles, so that we do not distinguish neutrinos from antineutrinos.

%%%%%%%%%%%%%%%%%%%%%%%%
\begin{table}
\caption{Matrix elements for $\nu_4$ production, see also~\cite{Hannestad:1995rs, Fuller:2009zz} for notation and details.}
\label{Matrixelementsterile}
\begin{center}
\begin{tabular}{c c}
\hline
{Process}& $S|{\cal M}|^2/(8G_F^2\sin^2\theta_{\tau 4})$\\
\hline
\hline
$\nu_{\tau}+\bar{\nu}_{\tau}\rightarrow\nu_4+\bar{\nu}_{\tau}$($\nu_{\tau}$)\quad\quad\quad\quad&$4u(u-m_4^2)$\\
$\nu_{\mu}+\bar{\nu}_{\mu}\rightarrow\nu_4+\bar{\nu}_{\tau}$($\nu_{\tau}$)\quad\quad\quad\quad&$u(u-m_4^2)$\\
$\nu_{\tau}+\nu_{\tau}\rightarrow\nu_4+\nu_{\tau}$\quad\quad\quad\quad&$2s(s-m_4^2)$\\
$\bar{\nu}_{\tau}+\bar{\nu}_{\tau}\rightarrow\nu_4+\bar{\nu}_{\tau}$\quad\quad\quad\quad&$2s(s-m_4^2)$\\
$\nu_{\mu}+\nu_{\tau}\rightarrow\nu_4+\nu_{\mu}$\quad\quad\quad\quad&$s(s-m_4^2)$\\
$\bar{\nu}_{\mu}+\bar{\nu}_{\tau}\rightarrow\nu_4+\bar{\nu}_{\mu}$\quad\quad\quad\quad&$s(s-m_4^2)$\\
$\nu_{\tau}+\bar{\nu}_{\mu}\rightarrow\nu_4+\bar{\nu}_{\mu}$\quad\quad\quad\quad&$u(u-m_4^2)$\\
$\bar{\nu}_{\tau}+\nu_{\mu}\rightarrow\nu_4+\nu_{\mu}$\quad\quad\quad\quad&$u(u-m_4^2)$\\
\hline
\label{Mandelstam Table}
\end{tabular}
\end{center}
\end{table} 
%%%%%%%%%%%%%%%%%%%%%%%%%%%%%%%%%%%%%%

%%%%%%%%%%%%%%%%%%%%%%%%%%%%%%%%%
\subsection{Kinetic equation for heavy sterile neutrino production}
%%%%%%%%%%%%%%%%%%%%%%%%%%%%%%%%%%%%%%%%

In order to characterize the heavy sterile neutrino emissivity one should solve the 
Boltzmann neutrino transport equation, as currently done for active neutrinos~\cite{Fischer:2009af}. This is a formidable process involving space-time evolution. In order to simplify 
this problem we assume the SN core as an homogeneous and isotropic environment, as in~\cite{Raffelt:1992bs}.
In this simplified situation, the equation governing the sterile neutrino production is~\cite{Hannestad:1995rs}:
%........................................
 \begin{equation}
 \frac{\partial f_s}{\partial t}={\mathcal C}_{\mathrm{coll}}(f) \,\ ,
 %...........................................
 \label{Boltzmann-equation}
 \end{equation}
where $f_s$ is the distribution of sterile neutrinos and ${\mathcal C}_{\mathrm{coll}}$ is the sum of all possible collisional interactions reported in Table~\ref{Matrixelementsterile}. We can express the collisional term as~\cite{Hannestad:1995rs}:
 \begin{equation}
  {\mathcal C}_{\mathrm{coll}}=\frac{1}{2E_s}\int d^3\hat{p}_2d^3\hat{p}_3d^3\hat{p}_4\Lambda (f_s,f_2,f_3,f_4)S|M|^2_{12\rightarrow 34} \delta^4(p_s+p_2-p_3-p_4)(2\pi)^4 \,\ ,
  \label{eq:Coll}
  \end{equation} 
where ${\rm d}^3\hat{p}= {\rm d}^3p/[(2\pi)^3 2E]$,
 $E_s$ and $p_s$ are energy and momentum of sterile neutrino respectively, 
$f_i$ is the distribution function of $i$-th neutrino, $S$ is a symmetry factor of $1/2$ for each couple of identical particles in the initial or final state, $|M|^2_{12\rightarrow34}$ refers to the processes in Table~\ref{Matrixelementsterile}. The function 
 %.................................
 \begin{equation}
\Lambda(f_1,f_2,f_3,f_4)=(1-f_1)(1-f_2)f_3f_4-f_1f_2(1-f_3)(1-f_4)\,\ ,
\end{equation}
%.....................................
 is the phase space factor, including the Pauli-blocking factor $(1-f_i)$ for the final states.
 Once produced, sterile neutrino will escape almost freely from the SN core, to the contrary of the active species which are diffusively confined within. Thus, to evaluate the collisional integral we assume $f_s=0$. Moreover, since $\nu_{\mu}$ and $\nu_{\tau}$ are in thermal equilibrium in the SN core, we can describe them by Fermi-Dirac distribution:
 %............................
\begin{equation}
f_{\nu_x}=\frac{1}{e^{E/T}+1} \,\ ,
\end{equation}
%.............................
with the temperature profile shown in Fig. \ref{fig:Temp}.

\begin{figure}[!t]
\centering
\includegraphics[width=0.6\textwidth]{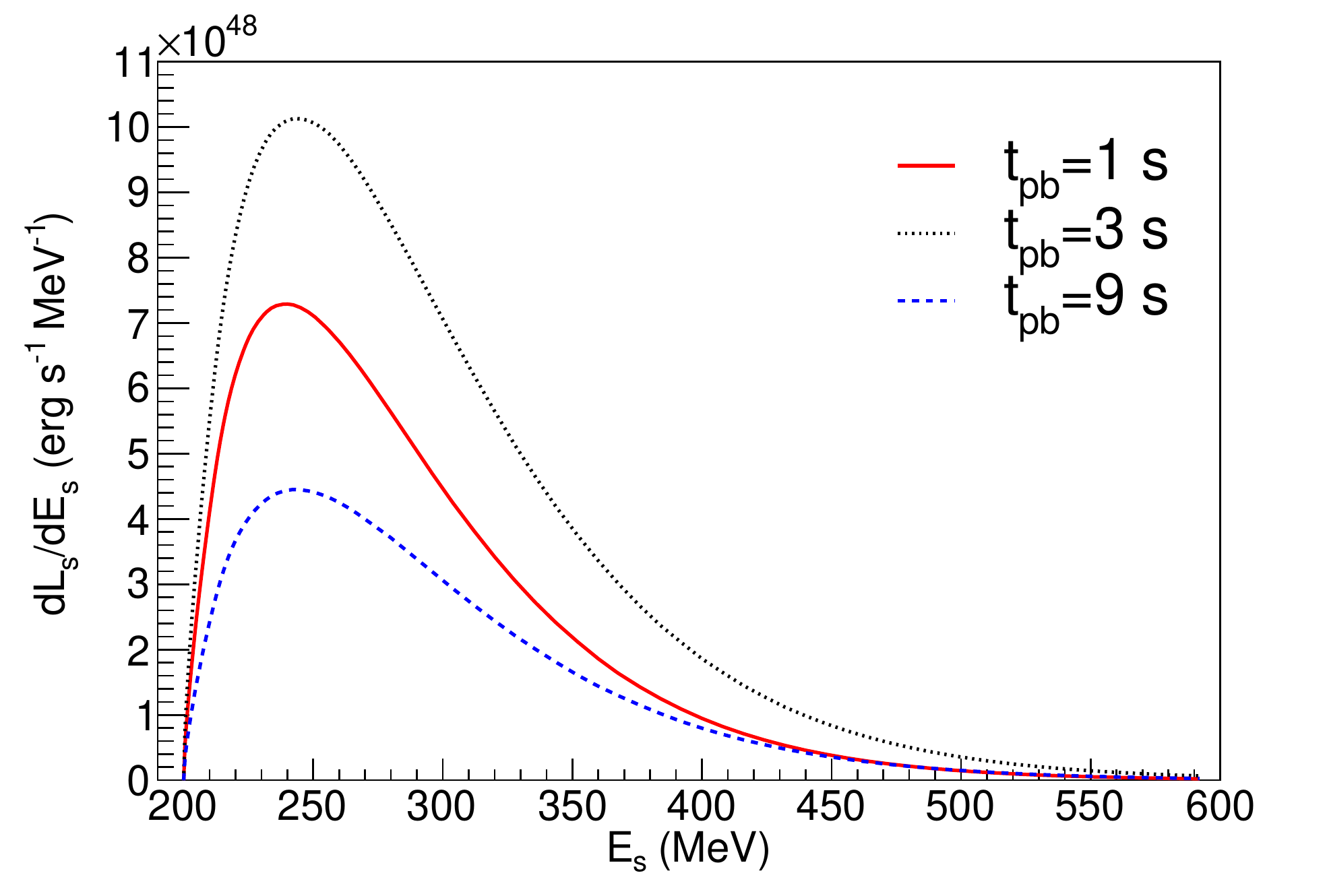}
\caption{Energy distribution of the $\nu_4$ luminosity at three different post-bounce times, for $m_4=200~\mathrm{MeV}$ and  $\sin^2\theta_{\tau 4}= 10^{-7}$.}
\label{luminosity03}
\end{figure}

\begin{figure}
\centering
\includegraphics[width=0.6\textwidth]{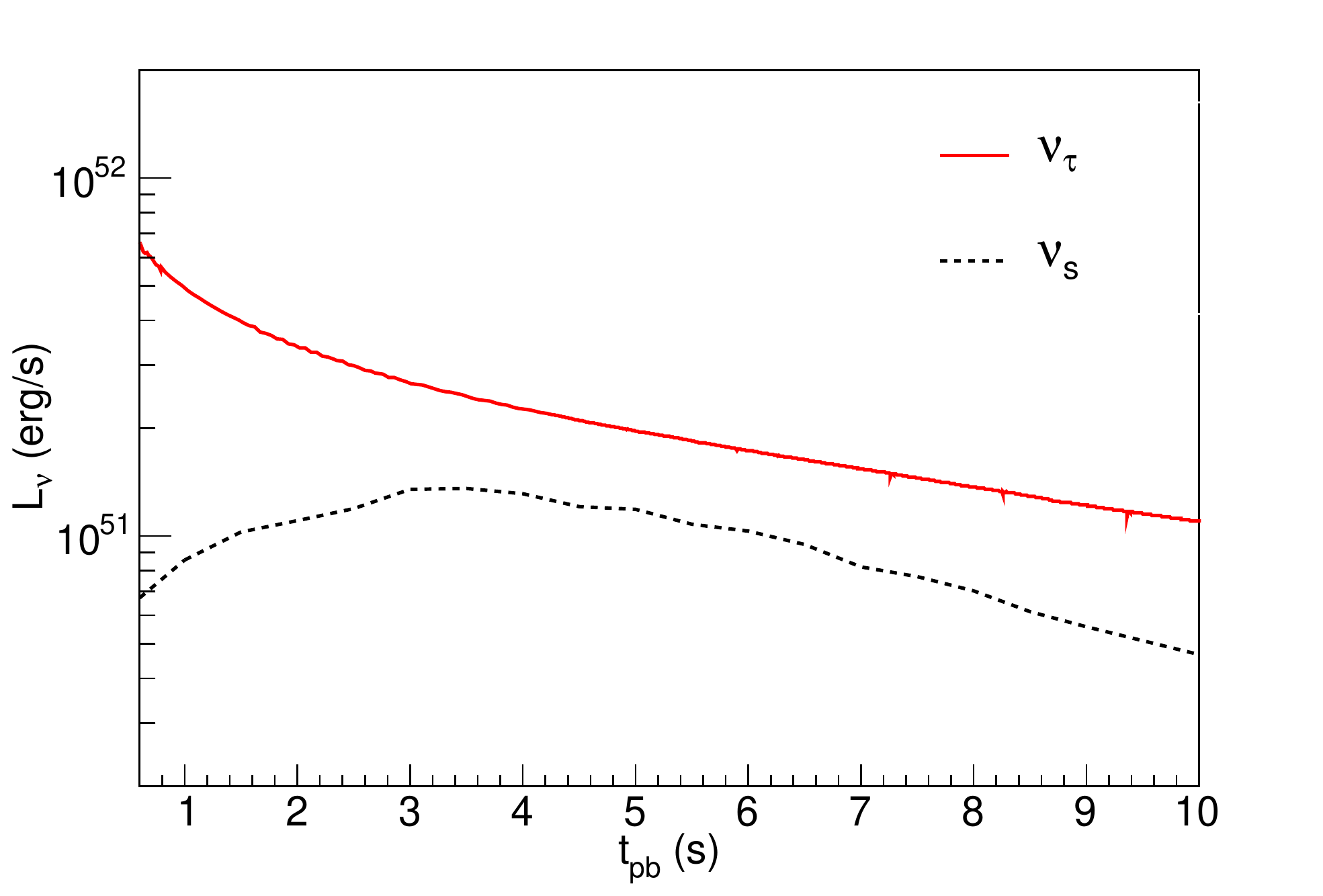}
\caption{Time-dependent comparison between the sterile neutrino luminosity and the $\nu_{\tau}$ luminosity emitted from our benchmark supernova model, assuming $m_4=200~\mathrm{MeV}$ and $\sin^2\theta_{\tau 4}= 10^{-7}$.}
\label{comparisonluminosity}
\end{figure}

We numerically solve the Boltzmann equation [Eq.~(\ref{Boltzmann-equation})] obtaining the sterile neutrino distribution  $f_s$. From this quantity, one can obtain
the sterile neutrino number density $n_s$ as~\cite{Tamborra:2017ubu}
 %..............................................
\begin{equation}
    {\rm d}n_s=\frac{f_s{\rm d}^3p}{(2\pi)^3}=\frac{f_sp_sE_s{\rm d}E_s{\rm d}\Omega }{(2\pi)^3} \,\ ,
    \label{dn(x)}
    \end{equation}
    %........................................
    so that 
%....................
\begin{equation}
\frac{{\rm d}n_s}{{\rm d}E_s}=\int {\rm d}\Omega\frac{f_sE_sp_s}{(2\pi)^3}=\frac{1}{2\pi^2} f_sE_sp_s \,\ .
\end{equation}
%........................................
Finally, integrating over the SN profile one gets the sterile neutrino energy spectrum  $d L_s/d E_s$  as 
%............................
\begin{align}
\frac{{\rm d}^2 N_s}{{\rm d}E_s{\rm d}t}&=\int {\rm d}^3r\frac{1}{2\pi^2} \frac{{\rm d}f_s}{{\rm d}t}E_sp_s=\frac{2}{\pi}\int {\rm d}rr^2 \frac{{\rm d}f_s}{{\rm d}t}E_sp_s \,\ \\
\frac{{\rm d}L_s}{{\rm d}E_s}&=E_s\frac{{\rm d}^2 N_s}{{\rm d}E_s{\rm d}t} \,\ .
\label{luminositysterile2}
\end{align}
%......................
In Fig.~\ref{luminosity03} we show the sterile neutrino energy spectrum ${\rm d}L_s/{\rm d}E_s$ at different post-bounce times for $m_4=200~\mathrm{MeV}$ and $\sin^2\theta_{\tau 4}=10^{-7}$. 
 One can see that the sterile neutrino energy spectrum is peaked at much higher energies with respect to the active neutrino ones. 
 As we will see, this peculiar feature provides a possible fingerprint of this flux.
Notice how  the sterile neutrino luminosity increases with time when $t_{\rm pb} \lesssim 3~\mathrm{s}$, due to the rise of the core temperature, while decreases at later times   due to cooling of the SN core  (see Fig.~\ref{fig:Temp}).
It is worth stressing that, for the chosen parameters, the sterile neutrino luminosity  is always smaller than the $\nu_{\tau}$ luminosity, as manifest from Fig.~\ref{comparisonluminosity}. This is an important {\it a posteriori} check of the consistency of our calculation in this range. 
Finally, we report that---for the assumed mass and mixing parameters of the sterile neutrino---the time-integrated function ${{\rm d} N_s}/{{\rm d} p_s}$, related to the energy spectrum simply via 
%.................................
\begin{equation}
\frac{{\rm d} N_s}{{\rm d} E_s}=\frac{\sqrt{p_s^2 + m_s^2}}{p_s}\frac{{\rm d} N_s}{{\rm d} p_s}   \,\, ,
\end{equation}
%................
can be fitted with the spectral representation of Eq.~(\ref{fesuppp}), considered as function of $p_s$, 
with parameters $N_s = 8.3 \times 10^{54}$, $\alpha = 1.64$ and $\langle p_s \rangle = 121$~MeV.
For completeness we  compare our sterile neutrino  emissivity  with the result of~\cite{Fuller:2009zz}. For  $m_4=200$~MeV and 
$\sin^2 \theta_{\tau 4}= 5 \times 10^{-8}$
they quote  an emitted energy $10^{51}$~erg in the time window $t_{\rm pb} \in [1;5]$~s. In the same situation we obtain $E_s \simeq 2.6 \times 10^{51}$~erg.

\begin{figure}[!t]
\centering
\includegraphics[width=0.6\textwidth]{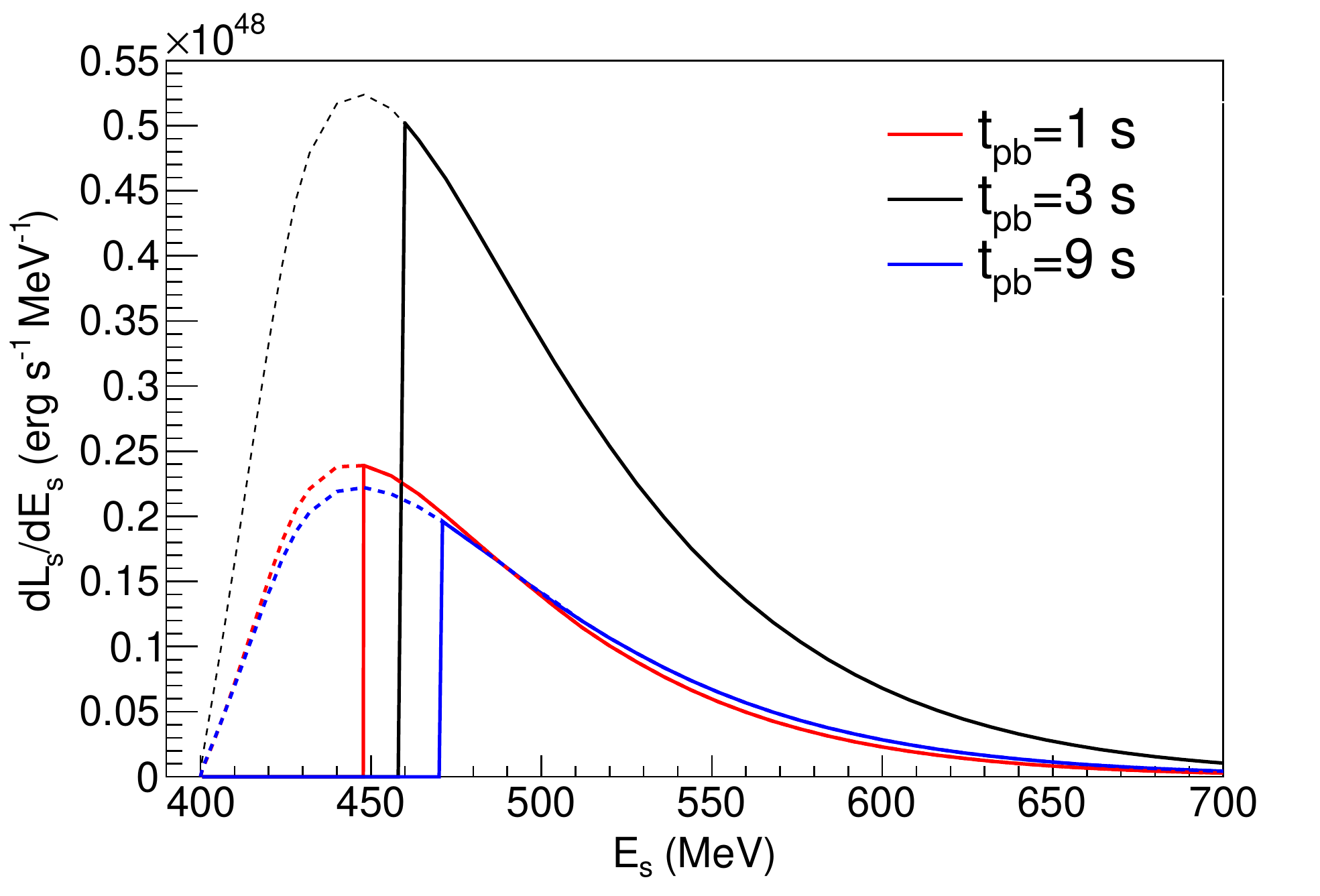}
\caption{Energy distribution of the $\nu_4$ luminosity at three different post-bounce times, for $m_4=400~\mathrm{MeV}$ and  $\sin^2\theta_{\tau 4}= 10^{-7}$.
Dashed lines refer to simulations without including the gravitational trapping effect, that is considered in the continuous curves.}
\label{luminosity03trapping}
\end{figure}

Finally, we mention that sterile neutrinos whose kinetic energy satisfies
%..........
\begin{equation}
E_{\rm kin} \leq K_{\rm tr} \equiv \frac{G_N M_r  m_4}{r} \,\ ,
\end{equation}
%............
will be trapped by gravitational attraction (see~\cite{Dreiner:2003wh} for details). In the above equation, $G_N$ is the Newton's constant, $r$ is the radius at which the sterile neutrino is produced, and $M_r$ is the mass of the supernova enclosed within the radius $r$. One can schematically include this effect by modifying the sterile neutrino energy spectrum as  
%.......................
\begin{equation}
\frac{d L_s}{d E_s} \to \frac{d L_s}{d E_s} \theta(E_s-K_{\rm tr} -m_4) \,\ .
\end{equation}
%.......................
We checked that for the case of $m_4=200$~MeV, shown in Fig.~\ref{luminosity03}, the effect of the trapping would only cut the energy spectrum for $E_s \lesssim 230$~MeV, producing a modest effect on the phenomenology discussed here. Thus, for simplicity, we neglected this ${\cal O}$(10\%) correction in the following. However, for higher masses the effect becomes sizable. In Fig.~\ref{luminosity03trapping}, we illustrate the case of $m_4= 400$~MeV and  $\sin^2\theta_{\tau 4}= 10^{-7}$. Dashed lines refer to simulations without including the gravitational trapping effect, that is considered in the continuous curves. We see that in the latter case about half of the energy spectrum is cut due to the gravitational trapping effect. Both the bounds and the phenomenology start to be seriously altered and the gravitational capture should not be neglected anymore.

%%%%%%%%%%%%%%%%%%%%%
%%%%%%%%%%%%%%%%%%%%%%%%
%%%%%%%%%%%%%%%%%%%%%
%%%%%%%%%%%%%%%%%%%%%%%%%%%%

%%%%%%%%%%%%%%%%%%%%%
%%%%%%%%%%%%%%%%%%%%%%%%

%%%%%%%%%%%%%%%%%%%%%%%%%%%%%%%
\subsection{SN 1987A bound on heavy sterile neutrinos}
%%%%%%%%%%%%%%%%%%%%%%%%%%%%%%%

\begin{figure}
\centering
\includegraphics[scale=0.6]{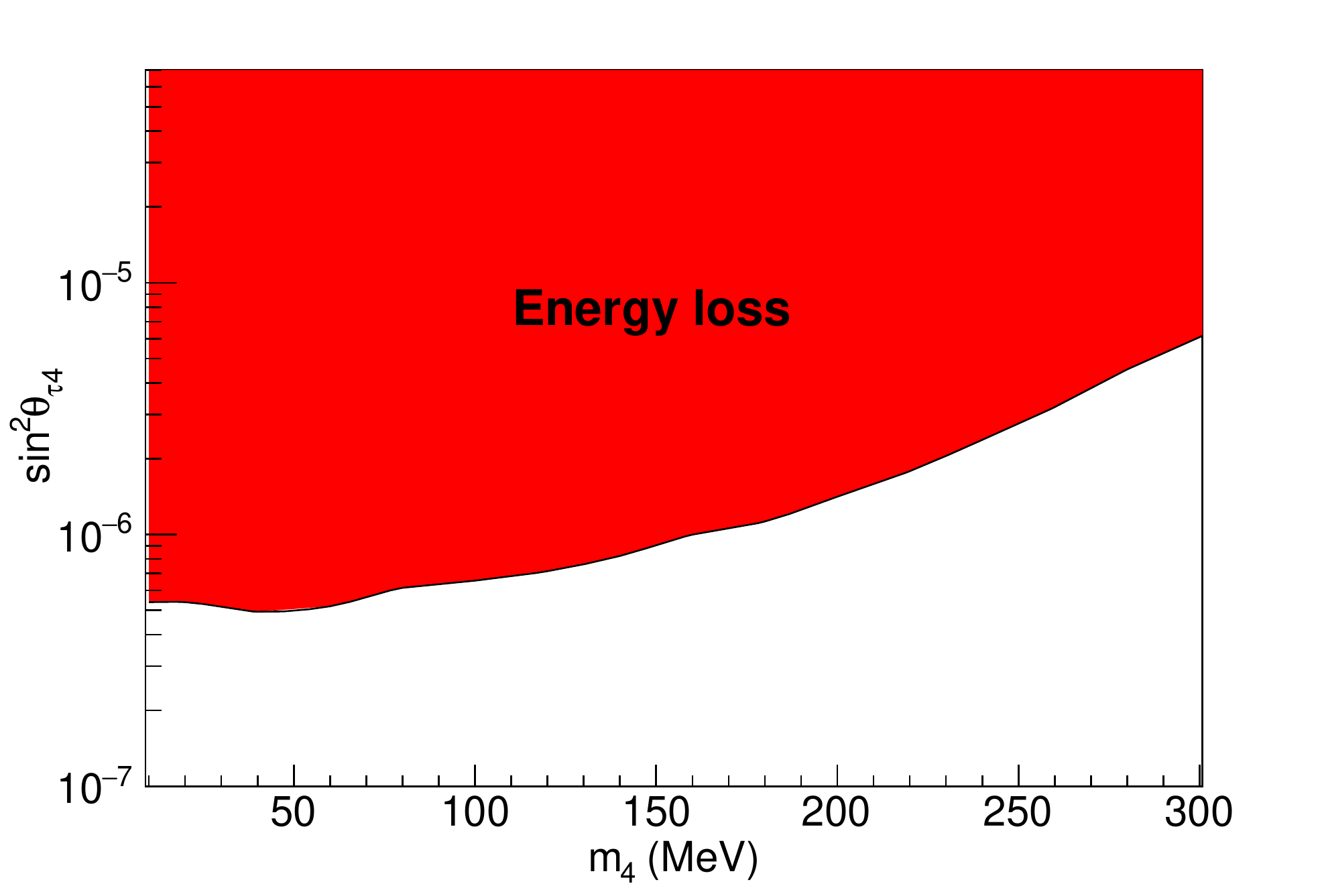}
\caption{The excluded region (in red) in the plane $(m_4,\sin^2\theta_{\tau 4})$ coming from the SN1987A energy loss argument applied to our benchmark SN model. The curve is the result of a smooth interpolation of bounds computed on a 10 MeV grid.}
\label{mitocca}
\end{figure}

Neutrino emission from SN 1987A can be used to obtain bounds on heavy sterile neutrinos~\cite{Raffelt:1987yt,Dolgov:2000pj,Dolgov:2000jw}. 
For small values of the mixing angle, like the ones we are considering, sterile neutrinos are weakly interacting in the SN core and freely escape after being produced, contributing to an additional energy-loss channel
on top of the standard one associated with active neutrinos.
In order to avoid conflict with the observations, one has to require the energy-loss rate per unit mass $\varepsilon_s\leq 1.0\times 10^{19}~\mathrm{erg~g^{-1}~s^{-1}}$\cite{Raffelt:2011nc} that can be translated into luminosity loss, considering the core mass $M\sim 1M_{\odot}$, via the relation
%%%%%%%%%%%%
\begin{equation}
L_s=\varepsilon_s\times 1\, M_{\odot}\simeq 2\times10^{52}\;\mathrm{erg/s} \,\ .
\end{equation}
%%%%%%%%%%%%%%%%%%%%
The red upper part of Fig.~\ref{mitocca} represents the region in $(m_4,\sin^2\theta_{\tau 4})$ nominally excluded by the 
energy-loss argument.~\footnote{We mention that a comparable bound has been 
obtained requiring that the energy transfer from heavy sterile neutrino decay into electromagnetic channels does not lead
to too energetic SN explosions~\cite{Rembiasz:2018lok,Sung:2019xie}).}
We stress that this part of the parameter space, in the case of a heavy sterile neutrino mixing with $\nu_\tau$, is unconstrained by laboratory experiments.
Note that for values of the mixing angle roughly above the showed region, the sterile neutrinos would be no longer free-streaming. They would enter a trapping regime, contributing to energy transfer in the SN core~\cite{Dolgov:2000pj,Dolgov:2000jw}. The study of this regime is more challenging, so that the bound cannot be naively extrapolated to large mixings. However, since large mixing angles are already excluded, it is not particularly interesting to investigate this situation from the phenomenological point of view. In the following, we will thus limit ourselves to consider mixing angles smaller than the ones excluded by SN 1987A. 
As discussed in the previous Section,  heavy sterile neutrinos with masses $m_4 \gtrsim 300$~MeV would be gravitationally trapped 
in the SN core, so the energy-loss bound would be significantly  relaxed at large masses.

%......................
%............................................
Note that the impact of  heavy sterile neutrinos on the SN 1987A, while typically overlooked, may have interesting applications. One recent example is the model~\cite{Gelmini:2019deq} in which  relatively heavy sterile neutrinos mixing with $\nu_\mu$ and $\nu_\tau$ have been argued to possibly relieve the tension between the ``locally measured'' vs. the ``cosmological inferred''  value of the Hubble parameter. 

%%%%%%%%%%%%%%%%%%%%%%%%%%%%%%
\section{Heavy sterile neutrino decays}\label{decayflux}
%%%%%%%%%%%%%%%%%%%%%%%%%%%%%
Heavy sterile neutrinos produced in SN core would decay in the stellar envelope producing an observable flux of active neutrinos. 
For sterile neutrinos with masses ${\mathcal O}(100)$~MeV and mixing only with $\nu_{\tau}$, the charged-current processes are kinematically forbidden. Thus, in the following we will only consider the allowed neutral-current decay
processes~\cite{Atre:2009rg,Bondarenko:2018ptm,Ballett:2019bgd,Bryman:2019bjg,Bryman:2019ssi}
%%%%%%%%%%%%%%%%%%%%%%%%%%%%%%%%%
\begin{eqnarray}
\nu_4&\rightarrow&  \nu\pi^0 \,\ , \nonumber   \\
\nu_4 &\rightarrow & \nu\nu\nu\,\ , \nonumber \\
\nu_4 &\rightarrow & \nu e^-e^+  \,\ ,
\label{eq:decays}
\end{eqnarray}
%%%%%%%%%%%%%%%%%%%%%%%%%%%%%%
which are the dominant ones in the mass range of interest. The latter is actually a factor four smaller than the three-neutrino one (in turn, sub-leading with respect to the two-body one) and will be eventually neglected, but to determine the branching ratios. Details on these processes are reported in the Appendix. For definitiveness, in the following numerical examples we will assume $m_4=200~\mathrm{MeV}$ and mixing $\sin^2\theta_{\tau 4}= 10^{-7}$.

%%%%%%%%%%%%%%%%%%%%%%%%
\subsection{Flux of daughter active neutrinos}
%%%%%%%%%%%%%%%%%%%%%%%%%%%%%
\label{Decayformula}

Consider the sterile neutrino decay with total rate $\Gamma$ and assume that some decay channel yields active neutrinos. The kinetic equations for the variation in the number of sterile $N_s(t^\prime)$ and active $N_a(t')$ neutrinos, due to the decay in the rest frame of the sterile neutrinos, are given by:
%%%%%%%%%%%%%%%%%%%%%%%%%%%%
\begin{align}
\frac{\rm d}{{\rm d}t^\prime}N_s(t')&=-\Gamma N_s(t')\,\ ,
\label{6} \\
\frac{\rm d}{{\rm d}t'}N_a(t')&=B_a\Gamma N_s(t') \,\ ,
\label{7}
\end{align}
%%%%%%%%%%%%%%%%%%%%%%%%%%%%
where $t'$ is the time in the rest frame of the sterile neutrino, and $B_a$ is the branching ratio of the considered decay process times the number of produced active neutrinos of type $a$. Passing from the rest frame to the laboratory frame, the general transformation of the active neutrino variables from primed quantities (referring to the sterile neutrino rest frame) to the unprimed (Lab) frame is:
%%%%%%%%%%%%%%%%%%%%
\begin{equation}
\left(
\begin{matrix}
E \\ p_{\perp} \\ p_{\parallel}
\end{matrix}
\right)
=
\left(
\begin{matrix}
\gamma& 0&  \beta\gamma \\
0 & 1 & 0\\
\beta\gamma& 0&  \gamma
\end{matrix}
\right)
\left(
\begin{matrix}
E'\\
p'\sin\theta\\
p'\cos\theta
\end{matrix}
\right)
\,\ ,
\label{trasformazioneCM}
\end{equation}
%%%%%%%%%%%%%%%%%%%%%%%
where $\gamma=1/\sqrt{1-\beta^2}$ is the Lorentz factor of the $\nu_4$ moving with velocity $\beta=v/c$ in the Lab frame, $\theta$ is the angle that the $\nu_a$ decay direction in the rest frame of $\nu_4$ forms with the velocity of $\nu_4$ in the Lab frame, and $p_\perp$
and $p_{\parallel}$ are the components of the $\nu_a$ momentum perpendicular and parallel to the $\nu_4$ momentum in the Lab frame, respectively.
From Eq.~(\ref{6}), one gets in the Lab frame
%%%%%%%%%%%%%%%%%%%%%%%%
\begin{equation}
\frac{\rm d}{{\rm d}t_D}N_s(t_D)=-\frac{1}{\tau\gamma}N_s(t_D) \Longrightarrow N_s(t_D)=e^{-t_D/\tau\gamma}N_s(0) \,\ ,
\label{10}
\end{equation}
%%%%%%%%%%%%%%%%%%%%
where and $\tau=1/\Gamma$ is the lifetime in the rest frame and and $t_D$ the time in the laboratory frame.
%%%%%%%%%%%%%%%%%%%%%%%%%
\begin{figure}
\centering
\includegraphics[scale=0.3]{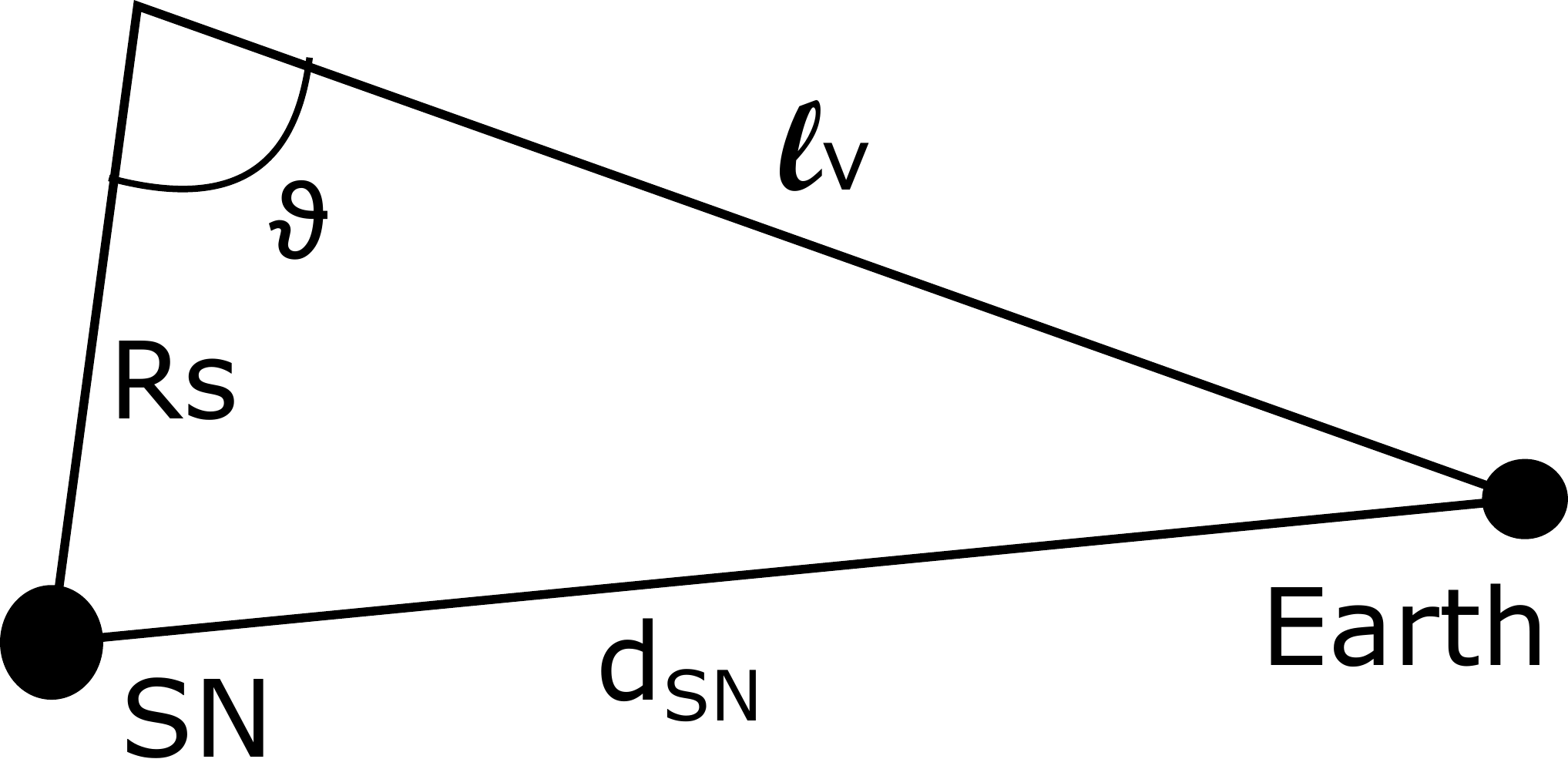}
\caption{Schematic representation (not on scale) of the path followed by a sterile neutrino and its daughter active neutrinos where $d_{\rm SN}$ is the distance between SN and Earth, $R_s$ is the distance travelled by the sterile before its decay and { $\ell_\nu$}  is the distance travelled by the daughter active neutrino to reach Earth.}
\label{path}
\end{figure}
%%%%%%%%%%%%%%%%%%%%%%%
The active neutrino emitted at a time $t_D$ inside the SN at a distance $R_s$ has to travel a distance  {$\ell_\nu$}  different from the Earth-SN center distance, $d_{\rm SN}$, as it is shown in Fig.~\ref{path}. Thus, the arrival time of the emitted active neutrino is $t_{\rm arr}=t_D+{\ell_{\nu}}/c$, to be compared with the minimum time required to arrive $t_{\rm min}=d_{\rm SN}/c$; we can choose the latter as zero, so that our time variable is $t=t_{\rm arr}-t_{\rm min}=t_D+({\ell_{\nu}}-d_{\rm SN})/c$ , i.e. the time delay. Using the law of cosines
%%%%%%%%%%%%%%%%%%%%%%%%%%%
\begin{equation}
d_{\rm SN}^2=R_s^2+{\ell_{\nu}}^2-2 {\ell_\nu}R_s\cos\mathcal{\vartheta} \,\ ,
\end{equation}
%%%%%%%%%%%%%%%%%%%%%%%%%%%
where $R_s$ is the distance from SN where the decay happens and $\mathcal{\vartheta}$ is the emission angle in the laboratory frame relative to the sterile neutrino momentum. For a decay inside  the SN with $R_s\ll d_{\rm SN}$, applicable to our situation of interest, we obtain
%%%%%%%%%%%%%%%%%%%%
\begin{equation}
{\ell_{\nu}}=d_{\rm SN}-R_s\cos\mathcal{\vartheta} \,\ .
\end{equation}
%%%%%%%%%%%%%%%%%%%%%%%%%%%
Therefore, since $\beta=R_s/(t_D\,c)$,
%%%%%%%%%%%%%%%%%%%%%%%
\begin{equation}
t=t_D(1-\beta\cos\mathcal{\vartheta})=\frac{t_D}{\gamma^2(1+\beta\cos\theta)} \,\ ,
\label{arrivaltime}
\end{equation}
%%%%%%%%%%%%%%%%%%%%%%
where we use the angular transformation
\begin{equation}
\cos\mathcal{\vartheta}=\frac{\beta+\cos\theta}{1+\beta\cos\theta} \,\ .
\end{equation}
Moreover,  using Eq.~(\ref{arrivaltime}) to express the time variable in terms of $t$, Eq.~(\ref{7}) can be rewritten in a differential form as~\cite{Oberauer:1993yr} 
\begin{equation}
\begin{split}
\frac{dN_a}{{\rm d}t{\rm d}E}=&\frac{B_a}{\tau}\int {\rm d}\cos\theta 
\int_{E_{\mathrm{min}}}^{\infty}{\rm d}E_s\frac{{\rm d}N_s(t,E_s)}{{\rm d}E_s}f_a\left(\frac{E}{\gamma(1+\beta\cos\theta)},\cos\theta\right)\\=&\frac{B_a}{\tau}\int {\rm d}\cos\theta\int_{E_{\mathrm{min}}}^{\infty}{\rm d}E_s\exp\Big[-\frac{\gamma(1+\beta\cos\theta)t}{\tau}\Big]\frac{{\rm d}N_s(0,E_s)}{dE_s} f_a\left(\frac{E}{\gamma(1+\beta\cos\theta)},\cos\theta\right) \,\ . 
\end{split}
\label{decaynumberconBa}
\end{equation}
%%%%%%%%%%%%%%%%%%%%%%%%%%%%%
where $f_a(E',\cos\theta)$ is the double differential distribution function (with respect to energy and angular variables) of the daughter active neutrino in the rest frame of $\nu_4$, normalized to 1 once integrated over $E'$ and $\cos\theta$. We have considered that for a massless active neutrino $E=p$, and $E_{\mathrm{min}}$ is the minimum sterile neutrino energy in the laboratory frame to produce an active neutrino with energy $E$. By integrating over the arrival delay time, one has
%%%%%%%%%%%%%%%%%%
\begin{equation}
\begin{split}
\frac{dN_a}{{\rm d}E}=B_a\int {\rm d}\cos\theta\int_{E_{\mathrm{min}}}^{\infty}{\rm d}E_s\,\frac{1}{\gamma(1+\beta\cos\theta)}\frac{{\rm d}N_s(0,E_s)}{dE_s} f_a\left(\frac{E}{\gamma(1+\beta\cos\theta)},\cos\theta\right)~, 
\end{split}
\label{decaytintegrated}
\end{equation}
%%%%%%%%%%%%%%%%%%%%%%%%%%%%%
where $\gamma=E_s/m_4$ and $\beta=p_s/E_s$. The neutrino yields corresponding to the dominant two and three body decay modes are described in detail in the Appendix, where this formula is applied to the concrete cases of interest.

\begin{figure}
\centering
\includegraphics[scale=0.6]{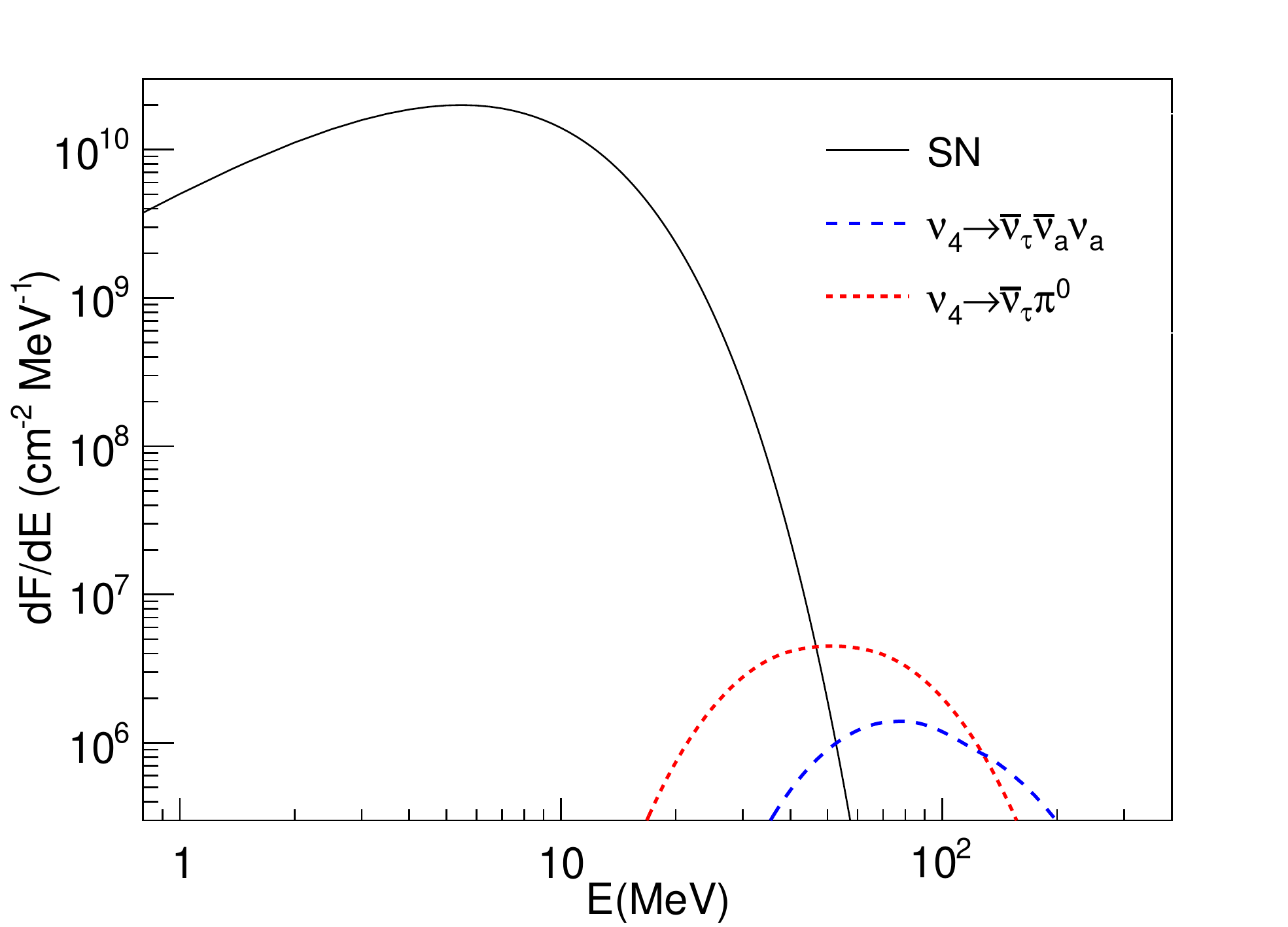}
\caption{$\nu_{\tau}$ energy distribution from SN explosion (solid curve),  $\nu_4\rightarrow\pi^0\nu_{\tau}$ (red, short-dashed curve) and $\nu_4\rightarrow\nu_{\tau}\nu_{\mu}\bar{\nu}_{\mu}$ decay (blue, long-dashed curve).}
\label{tauoni}
\end{figure}

In Fig.~\ref{tauoni} we compare the distribution of $\nu_{\tau}$'s from SN explosion simulation (solid curve) with the $\nu_{\tau}$'s produced by the sterile neutrino decay (dashed curves). We can see that the active neutrinos produced from $\nu_4\rightarrow\pi^0\nu_{\tau}$ (red, short-dashed curve) start to be dominant at $E\sim 50~\mathrm{MeV}$, while those produced from $\nu_4\rightarrow\nu_{\tau}\bar{\nu}_a\nu_a$ (blue, long-dashed) start to dominate at  $E\sim 120~\mathrm{MeV}$.
%%%%%%%%%%%%%%%%%%%%%%

%%%%%%%%%%%%%%%%%%%%%%%%%%%%%%%%%%
\section{Detecting the active neutrinos from sterile neutrino decays}\label{detection}
%%%%%%%%%%%%%%%%%%%%%%%%%%%%%%%%
In the following, we estimate the potential of large underground neutrino experiments to detect a neutrino burst induced by decaying heavy sterile neutrinos from SNe. For definiteness, we will focus on the underground water Cherenkov detector Super-Kamiokande (SK), with a fiducial mass of $22.5~\mathrm{kton}$, where the main detection channel for SN neutrinos is by
 inverse beta decay process $\bar{\nu}_e+p^+\rightarrow e^++n$. Estimates for the future Hyper-Kamiokande detector, whose 1TankHD design consists of $187~\mathrm{kton}$~\cite{Abe:2018uyc}, can be obtained by roughly rescaling the numbers quoted for SK by a factor 8.4.

%%%%%%%%%%%%%%%%%%%%%%%%%%%%%%%%
\subsection{Oscillated neutrino fluxes}
%%%%%%%%%%%%%%%%%%%%%%%%%%%%%%%%%%%%
\label{oscillatedneutrinoflux}

%%%%%%%%%%%%%%
The initial neutrino distributions are in general modified by flavor conversions $F^0_\nu \to F_\nu$. We assume a standard $3\nu$ framework where the mass spectrum of neutrinos is parameterized in terms of two mass-squared differences, whose
values are obtained from a $3\nu$ global analysis of the neutrino data~\cite{Capozzi:2018ubv}
\begin{eqnarray}
\left|\Delta m^2_{\rm atm}\right| &=& \left|m_3^2 -m_{1,2}^2\right| = 2.45 \times 10^{-3}~ \textrm{eV}^2~,\\
\Delta m^2_{\rm \odot} &=&  m_2^2 -m_1^2 = 7.34 \times 10^{-5}~\textrm{eV}^2~,
\end{eqnarray}
where according to the sign of $\Delta m^2_{\rm atm}$ one distinguishes a normal (NH,  $\Delta m^2_{\rm atm}>0$) or an inverted (IH,  $\Delta m^2_{\rm atm}<0$) mass ordering. The flavor eigenstates $\nu_{e,\mu,\tau}$ are linear combinations of the mass eigenstates $\nu_{1,2,3}$ with a unitary transformation parametrized by three mixing angles (the CP-violating phase do not enter in our calculations). Their best-fit values  of the relevant mixing angles, according to the global analysis, are (for the NH case)
\begin{eqnarray}
\sin^2 \theta_{12} = 0.304~, \;\;\; \sin^2 \theta_{13} = 0.0214~.
\end{eqnarray}
The value of the mixing angle $\theta_{23}$ is not relevant in our context since we will consider only the $\bar\nu_e$ detection. For IH case the best-fit values are similar to the ones quoted before.

\begin{figure}
\hspace{1.cm}
\includegraphics[width=0.5\textwidth]{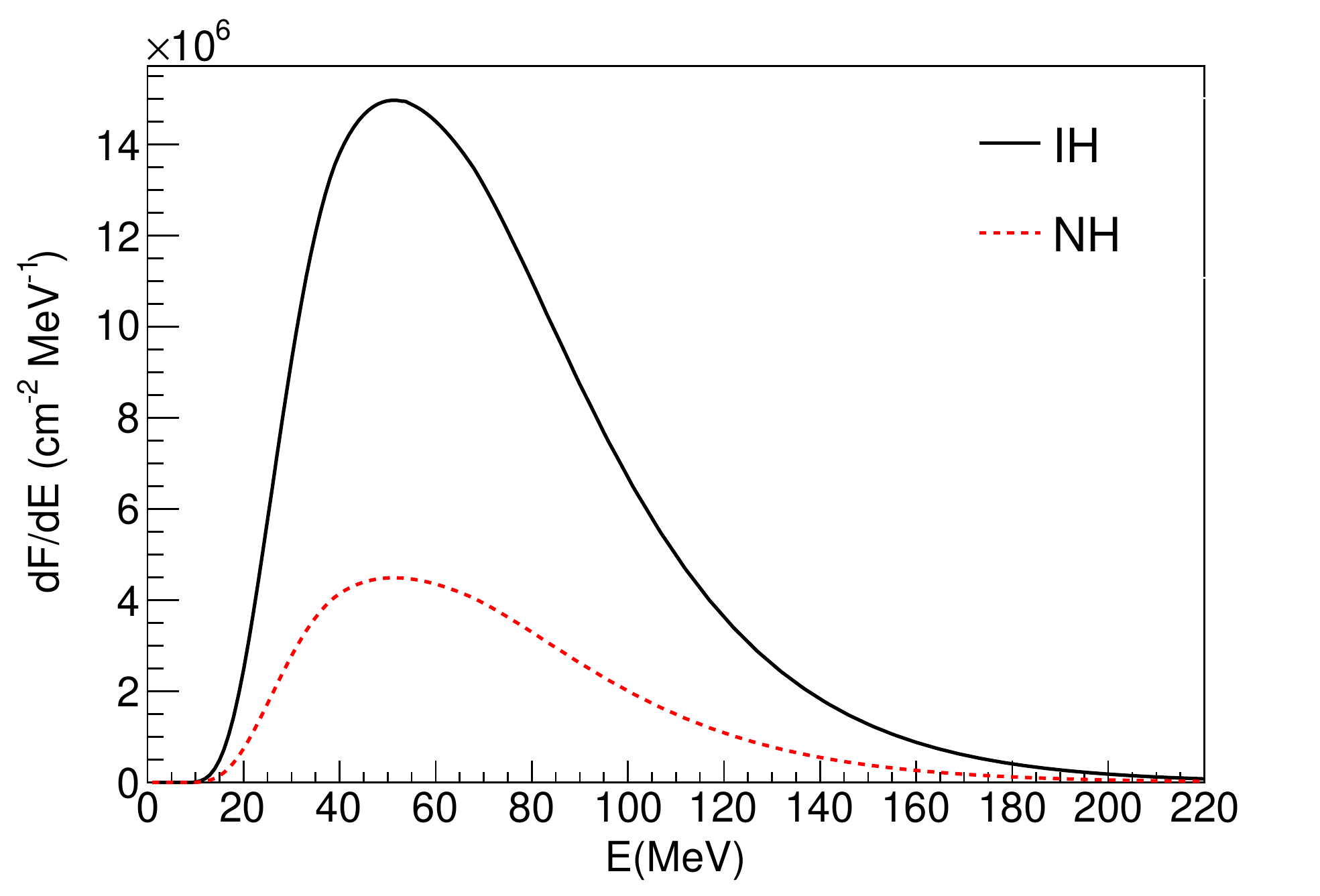}
\includegraphics[width=0.5\textwidth]{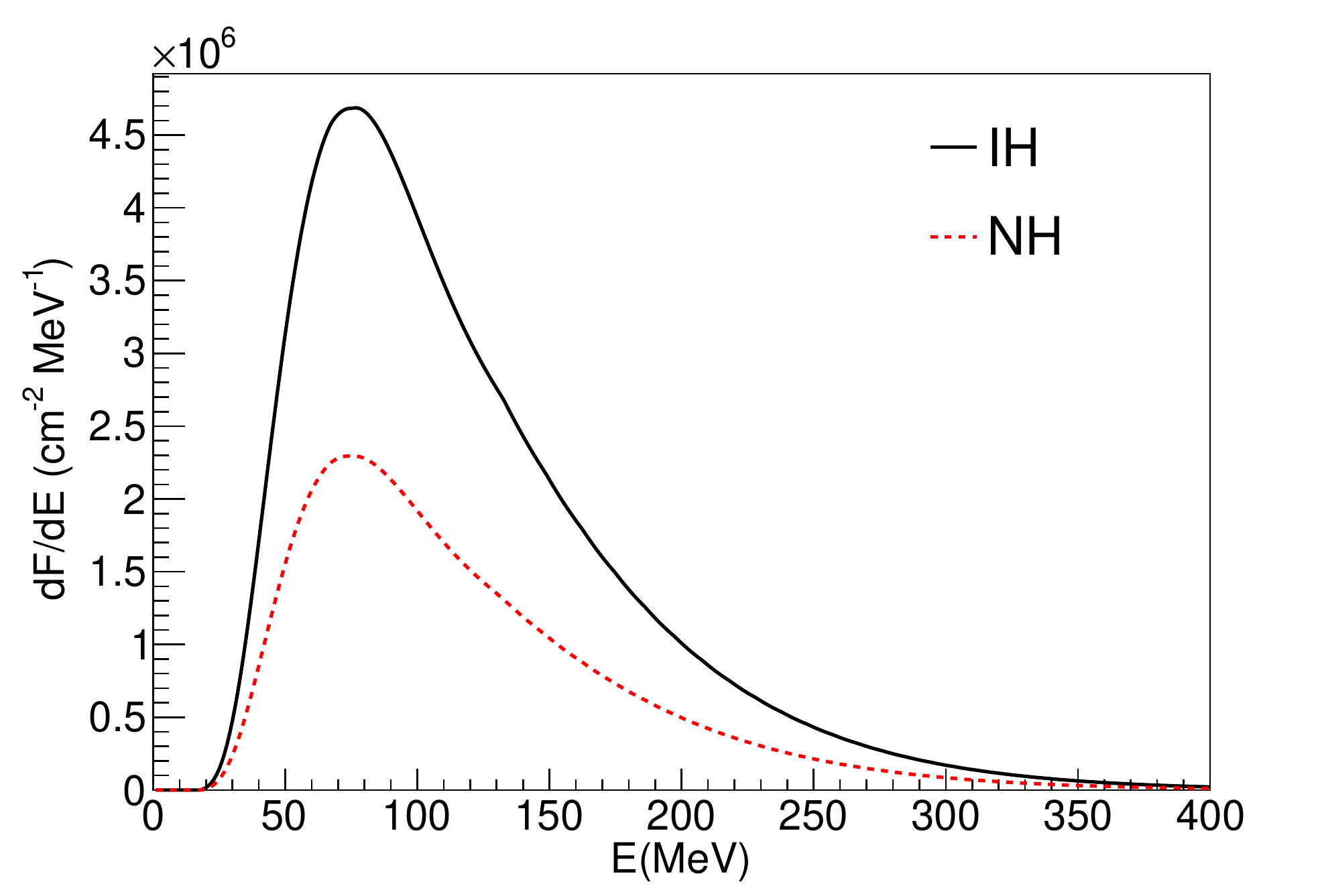}
\caption{$\bar{\nu}_e$ flux from heavy sterile neutrino decay  on Earth as a function of neutrino energy 
for a SN at a distance $d=10~\mathrm{kpc}$ for NH (dashed red curve) and IH (solid dark curve). Left panel:   $\nu_4\rightarrow\pi^0\nu_{\tau}$.
Right panel: $\nu_4\rightarrow\nu_{\tau}\nu_a\bar{\nu}_a$.}
\label{flux decayearth}
\end{figure}

Neutrino flavor conversions in SNe are a fascinating and complex phenomenon where different effects would contribute to profoundly modify the original neutrino fluxes (see Ref.~\cite{Mirizzi:2015eza} for a recent review). Indeed, in the deepest SN regions ($r \lesssim 10^3$~km) the neutrino density is sufficiently high to produce a self-induced refractive term for the neutrino propagation, associated with $\nu$--$\nu$ interactions. These would produce surprising collective effects in the flavor dynamics that are currently under investigation~\cite{Airen:2018nvp}. At larger radii ($r \sim 10^4$--$10^5$~km) neutrino fluxes would be further processed by the ordinary Mikheeyev-Smirnov-Wolfenstein (MSW)  matter effects~\cite{Wolfenstein:1977ue,Mikheev:1986gs,Dighe:1999bi}. In this context, 
shock-wave propagation and matter turbulences may induce additional modifications in the neutrino fluxes. 
Since our goal is not to give an accurate description of all these effects, in order to asses the detectability of the active neutrino signal induced by heavy sterile neutrino decays  
we adopt a schematic approach, assuming that  
the neutrino flux arriving at Earth can be expressed in terms of $\bar{\nu}_e$ energy-dependent survival probabilities $\bar{P}_{ee}(E)$~\cite{Mirizzi:2015eza}
%%%%%%%%%%%%%%%%
\begin{equation}
F_{\bar{\nu}_e}=\bar{P}_{ee}(E)F_{\bar{\nu}_e}^0+[1-\bar{P}_{ee}(E)]F_{\bar{\nu}_x}^0(E) \,\ .
\label{oscillationflux}
\end{equation}
%%%%%%%%%%%%%%%%%%%%
There is a similar equation for $F_{\nu_e}$ with probabilities $P_{ee}(E)$. We will use $\bar{P}_{ee}=\cos^2\theta_{12}$ for NH and $\bar{P}_{ee}=0$ for IH, where $\cos^2\theta_{12}\simeq 0.7$.
This choice corresponds to the sole matter effects along a static SN density profile. Using these equations we find the $\bar{\nu}_e$ flux from $\nu_4\rightarrow\pi^0\bar{\nu}_{\tau}$ and $\nu_4\rightarrow\nu_{\tau}\nu_a\bar{\nu}_a$ decays. In Fig.~\ref{flux decayearth}  we show for the NH (dashed curve) and IH (solid curve) $\bar{\nu}_e$ flux from $\nu_4\rightarrow\pi^0\bar{\nu}_{\tau}$ (left panel) and $\nu_4\rightarrow\nu_{\tau}\nu_a\bar{\nu}_a$ (right panel) decays, respectively. The fluxes at the Earth have been obtained considering the geometrical dilution factor $1/4\pi d^2$, where for definiteness here and in the following we assume a typical SN distance $d=10~\mathrm{kpc}$ (see e.g.~\cite{Mirizzi:2006xx}.)  
%%%%%%%%%%%%%%%%%%%%%%%%%%
%%%%%%%%%%%%%%%%%%%%%%%%%%%%%%%%

%%%%%%%%%%%%%%%%%%%%
\subsection{Events from sterile neutrino decay}
%%%%%%%%%%%%%%%%%

\begin{figure}
\centering
\includegraphics[scale=0.5]{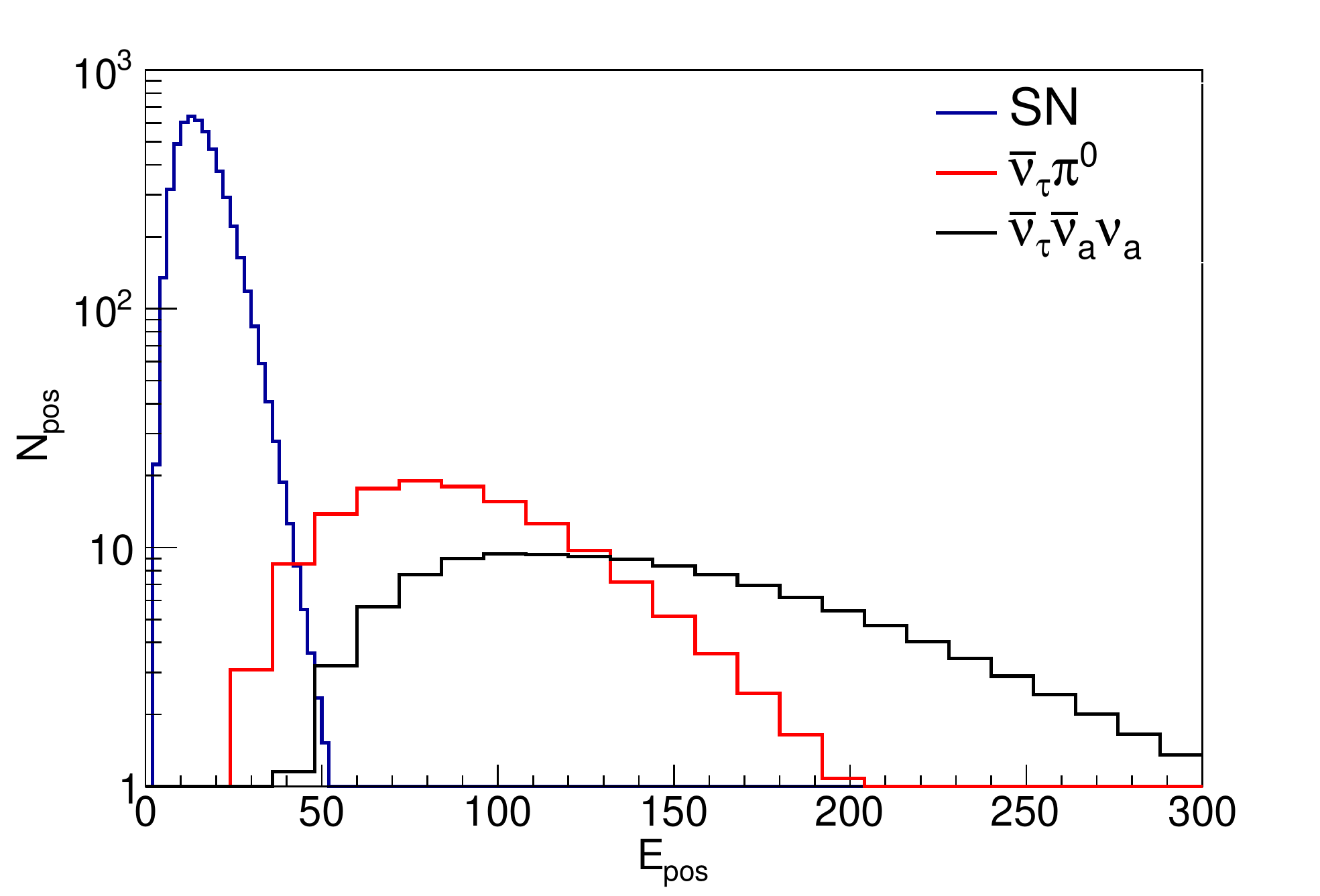}
\caption{Number of observable events in Super-Kamiokande as a function of the positron energy $E_{\rm pos}$, considering standard SN emission (blue, leftmost histogram), $\nu_4\rightarrow\bar{\nu}_{\tau}$ (red, middle histogram) and $\nu_4\rightarrow\nu_{\tau}\nu_a\bar{\nu}_a$ (black, rightmost histogram). NH is assumed.}
\label{confrontoeventiNH}
\end{figure}

In this section we characterize the signal in Super-Kamiokande, associated to the daughter neutrinos produced by heavy sterile neutrino decays.
Concerning the  inverse-beta-decay process, we take the differential cross section from Ref.~\cite{Strumia:2003zx}. The total cross section grows approximatively as $E^2$.
The number of events from the standard $\bar\nu_e$ signal and for the one induced by the decays are reported in Table~\ref{expected events} for both
neutrino mass hierarchies. For the chosen value of the mixing angle,  the sterile neutrino events constitute between 5\% and 10\% of the total flux. Nevertheless, the two contributions are separated in energy and this makes possible the detection of
the sterile induced signal. 
Indeed, in Fig.~\ref{confrontoeventiNH} we compare for NH the positron events energy distribution as function of positron energy in SK from SN explosion simulation (blue) with $\nu_4\rightarrow\pi^0\bar{\nu}_{\tau}$ (red) and $\nu_4\rightarrow\nu_{\tau}\bar{\nu}_a\nu_a$ (black) decays. The signal induced by heavy sterile $\nu$ decays is detectable since it would produce $\mathcal{O}(100)$ events in a region where in the standard scenario one does not expect a neutrino signal. 
Therefore,  the most distinctive and robust signature of the existence of $\nu_4$ would be a bump in the energy spectrum at $E_{\rm pos}\gtrsim 80~\mathrm{MeV}$.
Similar results would be obtained in IH.  Also, although with quantitative differences due to the different cross-sections and fiducial volumes, we expect this signature to be measurable at other detectors relying on techniques different from the Water Cherenkov one, provided that the `standard' SN signal generates at least a few hundreds events and that they have sufficient energy resolution.
%%%%%%%%%%%%%%%%%%%%%
\begin{table}
\caption{Expected events in SK from the `standard' SN $\bar{\nu}_e$ flux and those from $\nu_4$ decays.}
\vspace{0.5 cm}
\centering
\begin{tabular}{lcc}
\hline
 Channel & \multicolumn{2}{c}{Number of events}\\
 & NH &IH\\
  \hline
  \hline
SN $\bar{\nu}_e$&5280&5640\\  
$\nu_4\rightarrow\pi^0\bar{\nu}_{\tau}$&141&470\\
$\nu_4\rightarrow\nu_{\tau}\nu_a\bar{\nu}_a$&115&182\\
\hline
\end{tabular}
\label{expected events}
\end{table}
%%%%%%%%%%%%%%%%%%%%%%%%%%%%

%%%%%%%%%%%%%%%%%%%%%%%%%%%%%%%%
%%%%%%%%%%%%%%%%%%%%%%%%%%%%%%%%%%%%%%%

In principle, there is at least another signature of sterile neutrino decays, which is a certain time delay of the detectable active neutrino decay byproducts with respect to the active standard neutrino flux initially produced. Such a delay is not necessarily too small to be measurable with current timing capabilities, as for benchmark parameters the delay is in the tens of ms range (see Appendix~\ref{nudecayspectra}), possibly longer for smaller mixing angles or slightly lighter neutrinos. The major limitation in exploiting this information is in the lack of any specific time-stamp in the flux expected from the cooling phase. If the flux is really rather featureless as in the benchmark model used here, the timing information may turn out to be impossible to exploit. We point out, however, that a number of effects can imprint relatively fast timescales upon the neutrino signal, such as a rather abrupt termination of the cooling flux if  the proto-neutron star were to turn into a black hole over seconds timescales. Should the ``standard'', high-statistics neutrino flux present interesting time structures, the issue of searching for peculiar time-signatures in the high-energy flux attributable to sterile neutrino decays would certainly deserve to be revisited. 

Finally, we note that for a mixing with tau neutrinos, the flavour ratios of the high-energy flux from sterile decay are not identical to the conventional SN flux, even considering mixing. However, to exploit this information one would need a high statistics detection of the exotic flux in multiple channels (some sensitive to neutral currents), which would probably require multiple type of detectors and rather delicate combined analyses, and whose details strongly depend on the assumptions on the type of detectors available at the time.
 
%%%%%%%%%%%%%%%%%%%%%%%%%%%%
\section{Conclusions}\label{conclusions}
%%%%%%%%%%%%%%%%%%%%%%%%%%%%

Heavy sterile neutrinos with masses $\mathcal{O}(\mathrm{MeV-GeV})$
are predicted in extensions of the Standard Model such as  Neutrino Minimal Standard Model ($\nu$MSM).  Heavy sterile neutrinos in this mass range can be constrained with a series of laboratory experiments ranging from peak searches in meson decays to beam dumps. However, for the case of sterile $\nu$ mixing only with $\nu_{\tau}$ the laboratory constrains are quite weak. In this context, we showed that there is a region with mixing $\sin^2\theta_{\tau 4}\sim 10^{-7}$ and mass $m_s\sim 200~\mathrm{MeV}$ which is unconstrained by laboratory experiments, but where heavy sterile $\nu$ would be copiously emitted by a core collapse SN explosion, without violating the SN 1987A bound. The goal of this work has been to investigate this possibility and the opportunity to detect in large underground neutrino detectors signatures associated with the production of heavy sterile $\nu$ in SNe.\\
In order to characterize the sterile $\nu$ emissivity in SN, we solved the Boltzmann transport equation for the active-sterile neutrino system, characterizing the supernova environment through \textit{state-of-the-art} SN simulations.\\
These heavy sterile neutrinos produced in the supernova core would then decay in the stellar matter on a time scale of tens of ms. In particular, in this mass range the dominant decay processes are $\nu_4\rightarrow\nu_{\tau}\pi^0$ and $\nu_4\rightarrow\nu_{\tau}\nu_a\bar{\nu}_a$ where $a=e,\mu ,\tau$. For each of them we calculated the decay rate and the corresponding spectrum of the daughter active neutrinos.\\
Finally we characterized the possible signal due to daughter active neutrinos in the water Cherenkov detector Super-Kamiokande via inverse beta decay (IBD) $\bar{\nu}_e+p^+\rightarrow n+e^+$. We found that for the assumed mass and mixing parameters, sterile neutrinos would induce few hundred of events at $E>100~\mathrm{MeV}$ for a galactic SN at $d=10$~kpc where one does not expect a signal from supernova neutrinos in the standard scenario. Thus, the detection of a high-energy bump in the SN neutrino spectrum would be a clear signature of the sterile neutrino emission and cannot be mimicked by any other known process.
Of course one expects similar or better detection perspectives in future generation experiments like Hyper-Kamiokande~\cite{Abe:2018uyc},
 DUNE~\cite{Acciarri:2015uup} and JUNO~\cite{An:2015jdp}. Hyper-Kamiokande, for instance, would ensure a significant detection also for a SN further away in our Galactic neighborhood, for somewhat smaller mixing parameters/lighter sterile neutrino, or for a SN model with a core cooler than the one considered here by a few tens of percent.  
 
Although in this article we have not focused on the impact of the massive sterile neutrinos on the explosion mechanism, let us remind the reader that the sterile neutrino decay processes $\nu_4 \to \nu \pi^0$, would lead to a $\gamma$ flux from the  $\pi^0$ decay whose energetics is of the same order of the typical SN explosion energy (at least for our benchmark parameters). This photon injection would occur at distance ${\mathcal O}(10^3)$~km in the SN envelope. It has been pointed out in~\cite{Fuller:2009zz} that the energy deposition from the $\gamma$ flux would have an impact on the SN explosion, even if it occurs after the stalled shock-front. This effect has been numerically 
  studied in~\cite{Rembiasz:2018lok}, showing that the $\gamma$'s would heat the gas at higher radii that ejects matter from the outer core or the envelope while the center continues to accrete matter. So, it appears likely that for our choice of the  mass-mixing parameters of the sterile neutrinos one would turn models which are non-exploding in the standard 1D case
 into successful SN explosions, possibly with somewhat enhanced explosion energy. It is also probable that nucleosynthetic yields would differ in a heavy $\nu_s$-decay triggered SN or in an ordinary one. Needless to say, numerical simulations are needed to address these questions more quantitatively. Such a study may uncover further signatures or constraints.

The parameter space probed by a SN is largely within the reach of the future searches of heavy sterile neutrinos foreseen in the SHIP 
experiment~\cite{Alekhin:2015byh} or a wider range of collider searches (see~\cite{Chun:2019nwi} for a recent study), which would allow for an exciting laboratory counterpart of an eventual astroparticle signal. 
In conclusion, in addition to future laboratory searches, a Galactic supernova explosion would constitute a valuable opportunity to probe heavy sterile neutrinos such as the one predicted in the $\nu$MSM, and suitable searches for events beyond the energy range of typically predicted neutrino fluxes appear justified in the event of a SN neutrino burst.

%%%%%%%%%%%%%%%%%%
\section*{Acknowledgements}
%%%%%%%%%%%%%%%%%%%%%%%

We thank Tobias Fischer for providing the numerical data for the SN model we used in this work.
L.M. and A.M. warmly thank Pietro Colangelo and Fulvia De Fazio  for useful discussions during the development of this project. The work of A.M. is supported by the Italian Istituto Nazionale di Fisica Nucleare (INFN) through the ``Theoretical Astroparticle Physics'' project and by Ministero dell'Istruzione, Universit\`a e Ricerca (MIUR). The work of P.S. is supported by Univ. de Savoie via the AAP project ``DIGHESE''. A.E. thanks the partial support by the CNPq fellowship No.~310052/2016-5.

\appendix

%%%%%%%%%%%
\section{Relevant sterile neutrino decay modes}\label{nudecayspectra}
%%%%%%%%%%%
We consider both the leading two-body and three-body decays. In both cases we assume isotropically distributed decay products, as in~\cite{Oberauer:1993yr}: while this condition does not hold on an event-by-event basis, we deal with an ensemble of sterile neutrinos with random helicities, so that isotropy applies in a statistical sense. 
%%%%%%%%%%%%%%%%%%%%%%%%%%%%%%%%%%%%%%%%%%%
\subsection{Two-body decay $\nu_4\rightarrow\nu_{\tau}\pi^0$}\label{2bodyspectra}
%%%%%%%%%%%%%%%%%%%%%%%%%%%%%%%%%%%%%%%%%%%%%%
The decay rate of this process  is~\cite{Gorbunov:2007ak,Shrock:1982sc,Atre:2009rg,Bondarenko:2018ptm}
%...........................
\begin{equation}
\Gamma(\nu_4\rightarrow\pi^0+\nu_{\tau})%=\frac{\left(G_Ff_{\pi}\right)^2}{8\pi}\sin^2\theta_{\tau 4}~ m_4k^2
=\frac{\left(G_Ff_{\pi}\right)^2} {32\pi}\sin^2\theta_{\tau 4} ~m_4^3\left(1-\frac{m_{\pi}^2}{m_4^2}\right)^2 \,\ ,
\end{equation}
%.............................
where $f_\pi \simeq 135$~MeV is the pion decay constant.  The above formula is correct for the exclusive channel containing a neutrino final state, and is the same for Dirac and Majorana neutrinos. But since we assume a Majorana sterile neutrino, the charge conjugated process with a final antineutrino is also possible, so the total decay rate is twice the above, i.e.
\begin{equation}
\Gamma_{\rm 2\,body}=\Gamma(\nu_4\rightarrow\pi^0+\nu_{\tau})+\Gamma(\nu_4\rightarrow\pi^0+\bar\nu_{\tau})=2\Gamma(\nu_4\rightarrow\pi^0+\nu_{\tau})\,.
\end{equation}

The above rate, if compared to the total decay rate (see next section for the other channels), yields
a branching ratio $B_{\pi^0}=0.85$ at the reference mass $m_4=200\,$MeV.
As a consequence, the sterile neutrino lifetime $\tau=(\sum \Gamma_i)^{-1}$ is dominated by the two-body  process and we have  
%.....................
\begin{equation}
\begin{split}
\tau=\frac{1}{\Gamma_{\rm tot}}=\frac{B_{\pi^0}}{\Gamma_{\rm 2\,body}}\simeq &\,\, \frac{0.016~\mathrm{s}}{\left(1-0.46\left(\frac{200~\mathrm{MeV}}{m_4}\right)^2\right)^{2}}\times\left(\frac{10^{-7}}{\sin^2\theta_{\tau 4}}\right)^2\left(\frac{200~\mathrm{MeV}}{m_4}\right)^3 \,\ ,
\end{split}
\end{equation}
%..............................
from which we estimate the  sterile neutrinos decay length $d$:
\begin{equation}
\begin{split}
d=\tau \frac{p_s}{m_4}\simeq &\,\, \frac{3.6\times 10^3~\mathrm{km}}{\left(1-0.46\left(\frac{200~\mathrm{MeV}}{m_4}\right)^2\right)^{2}}\times\left(\frac{10^{-7}}{\sin^2\theta_{\tau 4}}\right)^2\left(\frac{200~\mathrm{MeV}}{m_4}\right)^3 \,\ ,
\end{split}
\end{equation}
where the numerical estimates at the RHS assumes a mean energy $E_s\sim 250~\mathrm{MeV}$. Therefore, sterile neutrinos decay inside the supernova envelope.

\begin{figure}
\centering
\includegraphics[scale=0.5]{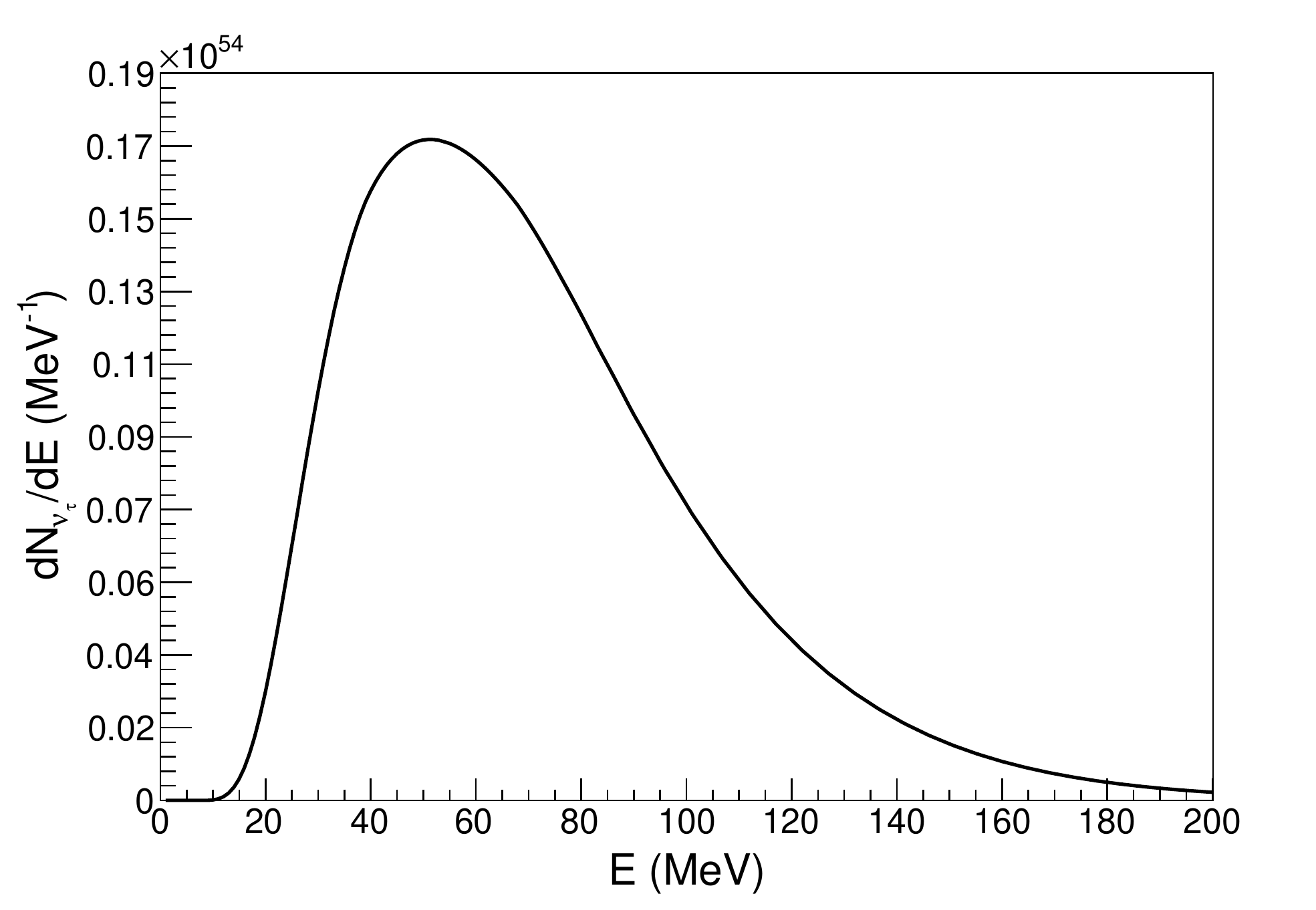}
\caption{The spectrum of $\nu_{\tau}$ from $\nu_4\rightarrow\nu_{\tau}\pi^0$ decay, for the benchmark SN model.}
\label{energydistributiondecay1}
\end{figure}

In this two-body process the daughter neutrino distribution in the decay frame, $f_{\nu_{\tau}}$, is  a Dirac delta in energy, and in the isotropic approximation one has
%%%%%%%%%%%%%%%%%%%
\begin{equation}
f_{\nu_{\tau}}=\frac{1}{2}\delta(E'-\overline{E}) \,\ ,
\end{equation}
%%%%%%%%%%%%%%%%%%%%%%%%
where
 %.........................
 \begin{equation}
 \overline{E}\equiv \frac{(m_4^2-m_{\pi}^2)}{2m_4}\;(\simeq 54\,{\rm MeV},\:{\rm for}\:m_4=200\,{\rm MeV})~,
 \end{equation}
 %..........................................
is the value that the energy of $\nu_{\tau}$ assumes in the center of mass of $\nu_4$.
To integrate  Eq.~(\ref{decaynumberconBa}) over ${\rm d}\cos\theta$, we perform a change of variable
\begin{equation}
E=\gamma(1+\beta\cos\theta)E'\rightarrow\cos\theta=\left(\frac{E}{\gamma E'}-1\right)\frac{1}{\beta} \,\ .
\label{EprimeE}
\end{equation}
Considering $E'$ as variable, we obtain:
%%%%%%%%%%%%
\begin{equation}
{\rm d}\cos\theta=-\frac{1}{\beta\gamma}\frac{E}{E'^{2}}{\rm d}E' \,\ .
\end{equation}
%%%%%%%%%%%%%%%%%%
From Eq.~(\ref{decaytintegrated}) and expressing $\gamma$ and $\beta$ as $E_s/m_4$ and $p_s/E_s$, respectively, keeping in mind that here a single neutrino is emitted per decay, we obtain
\begin{equation}
\frac{{\rm d}N_{\nu_{\tau}}}{{\rm d}E}=\frac{m_4}{2\overline{E}}B_{\pi^0}\int_{E_{\mathrm{min}}}^{\infty}{\rm d}E_s\frac{1}{p_s}\frac{{\rm d}N_s}{{\rm d}E_s} \,\ .
\label{formula flusso}
\end{equation}
Finally, it is possible to find $E_{\mathrm{min}}$ from Eq.~(\ref{EprimeE}) which we rewrite as \begin{equation}
E=\left(\frac{E_s+p_s\cos\theta}{m_4}\right)\overline{E} \,\ .
\end{equation}
 At fixed $E$, we obtain $E_{\mathrm{min}}$ when the factor in parentheses is maximized. This coincides with the request that $\cos\theta=1$. Thus, after a few algebraic steps one obtains
%%%%%%%%%%%%%
\begin{equation}
E_{\mathrm{min}}+\sqrt{E_{\mathrm{min}}^2-m_4^2}-m_4\frac{E}{\overline{E}}=0 \,\ ,
\end{equation}
%%%%%%%%%%%%%%
that leads to
%%%%%%%%%%%
\begin{equation}
E_{\mathrm{min}}=m_4\frac{E^2+\overline{E}^{2}}{2E\overline{E}} \,\ .
\label{Emindefinito}
\end{equation}
%%%%%%%%%%%%%%
In Fig.~\ref{energydistributiondecay1} we show the energy number distribution ${\rm d}N_{\nu_{\tau}}/{\rm d}E$ obtained from Eq.~(\ref{formula flusso}) and our benchmark SN model. The distribution  reaches a maximum at $E\sim60~\mathrm{MeV}$, significantly higher than the energy of the ordinary active neutrinos produced from SN explosion. Note also that since the sterile neutrinos are non-relativistic, the spectrum is essentially a thermal broadened distribution around the energy $\overline{E}$ (the zero-temperature limit would simply be a delta around $\overline{E}$).
%%%%%%%%%%%%%%%%%%%%%%%%%%
%%%%%%%%%%%%%%%%%%%%%%%

%%%%%%%%%%%%%%%%%%%%%%%%%%%%%%
\subsection{Three-body decays}\label{3bodyspectra}
%%%%%%%%%%%%%%%%%%%%%%%%%%%%%%%
\label{sterilethreebodydecay}
For sterile neutrinos mixing only with $\nu_\tau$,
the following processes are possible
%...................................
\begin{align}
\nu_4&\rightarrow \nu_{\tau}\nu_{a}\bar{\nu}_{a}, \,\bar{\nu}_{\tau}\nu_{a}\bar{\nu}_{a}\,\ ,
\label{Decaydacalcolaare}\\
\nu_4&\rightarrow \nu_{\tau}e^-e^+,\, {\bar\nu}_{\tau}e^-e^+ \,\ .
\label{Decaydanoncalcolare}
\end{align}
%.........................

Exclusive decay rates are equal for Dirac and Majorana cases. For the three neutrino case one has 
%%%%%%%%%%%%%%%%%%%%%%%%
\begin{equation}
\Gamma(\nu_4\rightarrow \nu_{\tau}\nu_{\beta}\bar{\nu}_{\beta})=\frac{G_F^2}{768\pi^3}\sin^2\theta_{\tau 4} m_4^5(1+\delta_{\tau\beta})\label{exclusive3nu} \,\ .
\end{equation}
%%%%%%%%%%%%%%%%%%%%%%%%
In the case of interest, i.e. a Majorana neutrino, one has to account for the charge conjugated processes to get the total rate~\cite{Gorbunov:2007ak,Shrock:1982sc,Atre:2009rg,Bondarenko:2018ptm}
\begin{equation}
\Gamma_{\rm 3}\equiv\sum_\beta[\Gamma(\nu_4\rightarrow \bar{\nu}_{\tau}\nu_{\beta}\bar{\nu}_{\beta})+\Gamma(\nu_4\rightarrow \nu_{\tau}\nu_{\beta}\bar{\nu}_{\beta})]=\frac{G_F^2}{96\pi^3}\sin^2\theta_{\tau 4} m_4^5 \label{total3nu}\,\ .
\end{equation}

Similarly, for the $e^\pm$ channel, assuming $m_e\ll m_4$~\cite{Gorbunov:2007ak,Shrock:1982sc,Atre:2009rg,Bondarenko:2018ptm}:
%%%%%%%%%%%%%%%%%%%%%%%%
\begin{equation}
\Gamma(\nu_4\rightarrow \nu_{\tau}e^+e^-)=\frac{G_F^2}{768\pi^3}\sin^2\theta_{\tau 4} m_4^5\left(1+4\sin^2\theta_W+8\sin^2\theta_W\right)~,
\end{equation}
%%%%%%%%%%%%%%%%%%%%%%%%
yielding the total decay width into an electron-positron pair
%%%%%%%%%%%%%%%%%%%%%%%%
\begin{equation}
\Gamma_ {e^\pm}\equiv \Gamma(\nu_4\rightarrow \nu_{\tau}e^+e^-)+\Gamma(\nu_4\rightarrow \bar{\nu}_{\tau}e^+e^-)=2 \Gamma(\nu_4\rightarrow \nu_{\tau}e^+e^-) \,\ .
\end{equation}
%%%%%%%%%%%%%%%%%%%%%%%%
Since at the reference mass of $m_4=200$ MeV this width is a factor of four smaller than the three-neutrino one, which in turn is seven times smaller than the two-body dominant width, or equivalently they have branching ratios $B_{e^\pm}:B_{3}:B_{\pi^0}\simeq 0.03:0.12:0.85$, we neglect the neutrinos coming from the $e^+e^-$ channel.

The double differential spectrum (in energy and angle) of the neutrinos in the sterile neutrino rest frame, normalized to one, is either
%%%%%%%%%%%%%%%%%%
\begin{equation}
f_I=\frac{1}{2}\frac{256}{27} \frac{E'^2}{m_4^3}\left(3-4\frac{E'}{m_4}\right) \,\ ,
\label{fI}
\end{equation}
%%%%%%%%%%%%%%%%%%%%%%%
or
%%%%%%%%%%%%%%%%%%%%%%%
\begin{equation}
f_{II}=\frac{1}{2}96 \frac{E'^2}{m_4^3}\left(1-2\frac{2E'}{m_4}\right)  \,\ .
\label{fII}
\end{equation}
%%%%%%%%%%%%%%%%%%%%%%%
The spectrum $f_I$ is the spectrum of both neutrinos in the decays $\nu_4\to\nu_\tau\nu_\beta\bar\nu_\beta$, while the spectrum $f_{II}$ is the spectrum of the antineutrino. For the charge-conjugated channel  $\nu_4\to\bar{\nu}_\tau\nu_\beta\bar\nu_\beta$, $f_I$ characterizes the antineutrinos, $f_{II}$ the neutrinos. These relations can be simply derived from the well-known case of the muon decay.

Compared with the two-body decay case, both the spectral distribution and the particle multiplicity are more involved to determine. 
Each process involved in Eq.~(\ref{exclusive3nu}) involves the emission of one $\nu_e$ with spectrum $f_I$ and one $\bar\nu_e$ with spectrum $f_{II}$, while its conjugate involves the emission of one $\nu_e$ with spectrum $f_{II}$ and one $\bar\nu_e$ with spectrum $f_I$; {\it mutatis mutandis}, the same holds for the $\mu$ flavour. On the other hand, all the $\nu_\tau$'s emitted in the $\nu_4\to\nu\nu\bar\nu$ process have distribution $f_I$, while the $\nu_\tau$ coming from the charge-conjugated one has distribution $f_{II}$. As a result, a simple counting shows that 
the differential rate of the all flavour species but for the third family is
\begin{equation}
\frac{{\rm d}\Gamma_{e}}{{\rm d}E'{\rm d}\cos\theta}=\frac{{\rm d}\Gamma_{\bar e}}{{\rm d}E'{\rm d}\cos\theta}=\frac{{\rm d}\Gamma_{\mu}}{{\rm d}E'{\rm d}\cos\theta}=\frac{{\rm d}\Gamma_{\bar \mu}}{{\rm d}E'{\rm d}\cos\theta}=\frac{\Gamma_3}{8}(f_I+f_{II})\,,
\end{equation}
%%%%%%%%%%%%%%%%%%
while for $\tau$ and $\bar\tau$ one has
%%%%%%%%%%%%%%%%%%
\begin{equation}
\frac{{\rm d}\Gamma_{\tau}}{{\rm d}E'{\rm d}\cos\theta}=\frac{{\rm d}\Gamma_{\bar\tau}}{{\rm d}E'{\rm d}\cos\theta}=\frac{\Gamma_3}{4}(3\,f_I+f_{II})\,.
\end{equation}
%%%%%%%%%%%%%%%%%%
Note that  the above expressions divided by  $\Gamma_{\rm tot}=\Gamma_{\rm 2\,body}+\Gamma_3+\Gamma_{e^\pm}$, factorizing out the normalized energy-angle distribution, can be used to isolate the pre-factor $B_i$ (branching ratio times multiplicity) symbolically appearing in Eq.~(\ref{decaytintegrated}), if one so wishes. 
We finally perform the same change of variable as for the two-body decay, Eq.~(\ref{EprimeE}), express $\gamma$ and $\beta$ as $E_s/m_4$ and $p_s/E_s$, respectively, 
and obtain similarly to the two-body case
%%%%%%%%%%%%%%%%%%%%%
\begin{equation}
\frac{{\rm d}N_{\alpha}}{{\rm d}E}=\int_{E_{\mathrm{min}}}^{\infty}{\rm d}E_s\int_{\cos\theta_{\mathrm{min}}}^{1}{\rm d}\cos\theta \frac{{\rm d}N_s}{{\rm d}E_s}\frac{1}{\Gamma_{\mathrm{tot}}} \frac{{\rm d}\Gamma_{\alpha}}{{\rm d}E{\rm d}\cos\theta}\,\ ,\label{ref1}
\end{equation}
%%%%%%%%%%%%%%%%%%%%%
where 
$\cos\theta_{\mathrm{min}}$ is determined by the maximum energy that a produced neutrino can have in the center of mass of the sterile one, and depends on the shape of $f_I$ or $f_{II}$.

\begin{figure}
\centering
\includegraphics[scale=0.6]{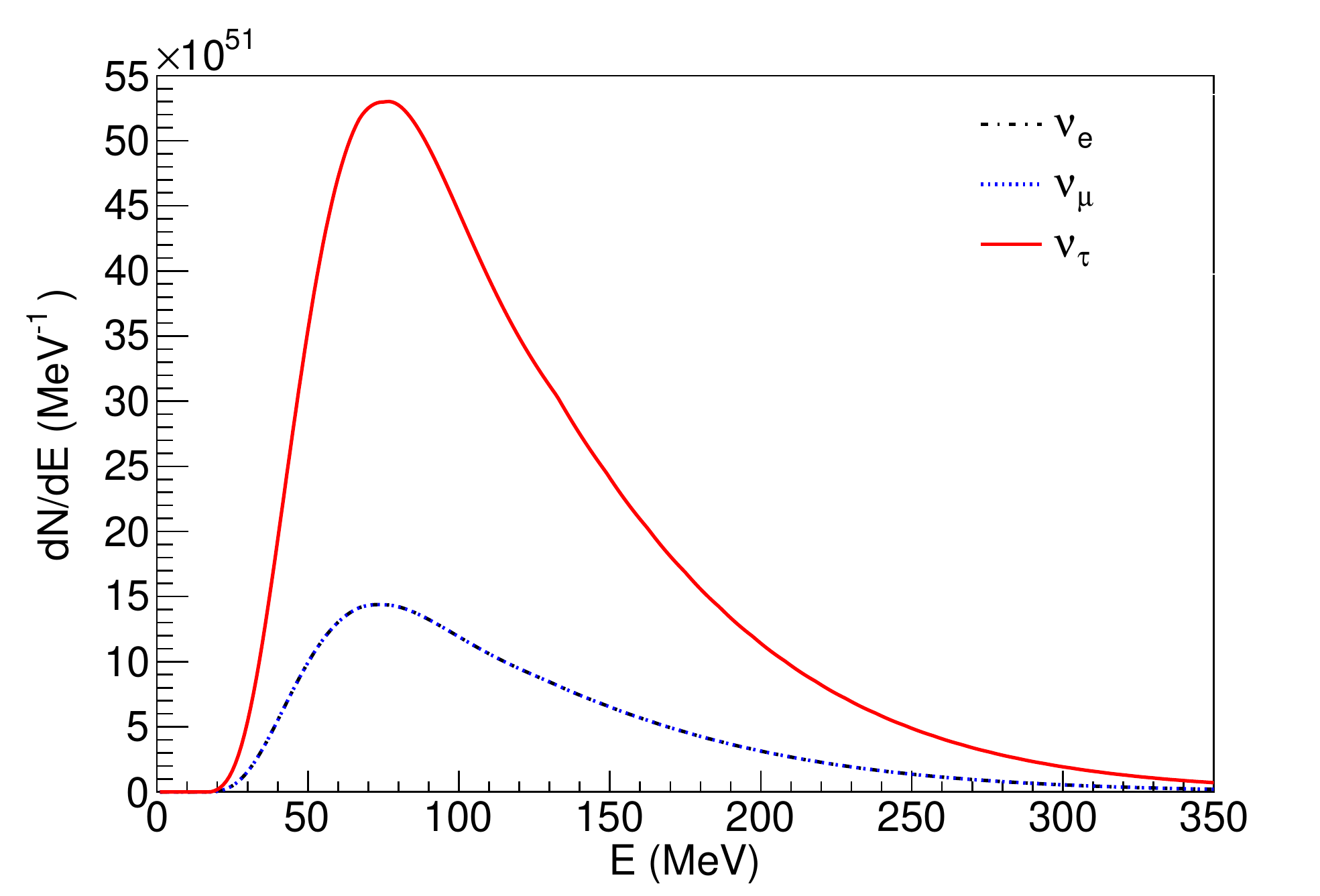}
\caption{Energy number distribution for $\nu_{i}$ with $i=e,\mu,\tau$ from the $\nu_4\rightarrow\nu_{\tau}\nu_i\bar{\nu}_i$ decay, for our benchmark SN.}
\label{energydistributiondecay2}
\end{figure}

By numerically integrating  Eq.~(\ref{ref1}) we obtain the $\nu_{e}$, $\nu_{\mu}$ and $\nu_{\tau}$ energy distributions ${\rm d}N_{\nu_a}/{\rm d}E$ (and their identical, charge-conjugated ones) from the three neutrino decay,  that we plot in Fig.~\ref{energydistributiondecay2}. The distribution reaches a maximum at {$E\sim80~\mathrm{MeV}$} and extends beyond the two-body decay one, since the three particles in the final states are massless. 

 %%%%%%%%%%%%%%%%%%%%%
%%%%%%%%%%%%%%%%%%%%%%%%%%%%%

%%%%%%%%%%%%%%%%%%%%%%%%%%%%%%%%%%%%%%%%%%%%%%%%%%%%%%%%%%%%%%%%%%%%%%
%\section*{References}
%%%%%%%%%%%%%%%%%%%%%%%%%%%%%%%%%%%%%%%%%%%%%%%%%%%%%%%%%%%%%%%%%%%%%%


\begin{thebibliography}{99}

%\cite{Mirizzi:2015eza}
\bibitem{Mirizzi:2015eza} 
  A.~Mirizzi, I.~Tamborra, H.~T.~Janka, N.~Saviano, K.~Scholberg, R.~Bollig, L.~Hudepohl and S.~Chakraborty,
  ``Supernova Neutrinos: Production, Oscillations and Detection,''
  Riv.\ Nuovo Cim.\  {\bf 39}, no. 1-2, 1 (2016)
%  doi:10.1393/ncr/i2016-10120-8
  [arXiv:1508.00785 [astro-ph.HE]].
  %%CITATION = doi:10.1393/ncr/i2016-10120-8;%%
  
  %\cite{Horiuchi:2017sku}
\bibitem{Horiuchi:2017sku} 
  S.~Horiuchi and J.~P.~Kneller,
  ``What can be learned from a future supernova neutrino detection?,''
  J.\ Phys.\ G {\bf 45}, no. 4, 043002 (2018)
%  doi:10.1088/1361-6471/aaa90a
  [arXiv:1709.01515 [astro-ph.HE]].
  %%CITATION = doi:10.1088/1361-6471/aaa90a;%%
  
  %\cite{Raffelt:2012kt}
\bibitem{Raffelt:2012kt} 
  G.~G.~Raffelt,
  ``Neutrinos and the stars,''
  Proc.\ Int.\ Sch.\ Phys.\ Fermi {\bf 182}, 61 (2012)
 % doi:10.3254/978-1-61499-173-1-61
  [arXiv:1201.1637 [astro-ph.SR]].
  %%CITATION = doi:10.3254/978-1-61499-173-1-61;%%
  
  %\cite{Carenza:2019pxu}
\bibitem{Carenza:2019pxu} 
  P.~Carenza, T.~Fischer, M.~Giannotti, G.~Guo, G.~Martinez-Pinedo and A.~Mirizzi,
  ``Improved axion emissivity from a supernova via nucleon-nucleon bremsstrahlung,''
  arXiv:1906.11844 [hep-ph].
  %%CITATION = ARXIV:1906.11844;%%
  
  %\cite{Chang:2016ntp}
\bibitem{Chang:2016ntp} 
  J.~H.~Chang, R.~Essig and S.~D.~McDermott,
  ``Revisiting Supernova 1987A Constraints on Dark Photons,''
  JHEP {\bf 1701}, 107 (2017)
%  doi:10.1007/JHEP01(2017)107
  [arXiv:1611.03864 [hep-ph]].
  %%CITATION = doi:10.1007/JHEP01(2017)107;%%
  
  %\cite{Hannestad:2001jv}
\bibitem{Hannestad:2001jv} 
  S.~Hannestad and G.~Raffelt,
  ``New supernova limit on large extra dimensions,''
  Phys.\ Rev.\ Lett.\  {\bf 87}, 051301 (2001)
%  doi:10.1103/PhysRevLett.87.051301
  [hep-ph/0103201].
  %%CITATION = doi:10.1103/PhysRevLett.87.051301;%%
  
  %\cite{Hannestad:2007ys}
\bibitem{Hannestad:2007ys} 
  S.~Hannestad, G.~Raffelt and Y.~Y.~Y.~Wong,
  ``Unparticle constraints from SN 1987A,''
  Phys.\ Rev.\ D {\bf 76}, 121701 (2007)
%  doi:10.1103/PhysRevD.76.121701
  [arXiv:0708.1404 [hep-ph]].
  %%CITATION = doi:10.1103/PhysRevD.76.121701;%%
  
  %\cite{Raffelt:2011nc}
\bibitem{Raffelt:2011nc} 
  G.~G.~Raffelt and S.~Zhou,
  ``Supernova bound on keV-mass sterile neutrinos reexamined,''
  Phys.\ Rev.\ D {\bf 83}, 093014 (2011)
%  doi:10.1103/PhysRevD.83.093014
  [arXiv:1102.5124 [hep-ph]].
  %%CITATION = doi:10.1103/PhysRevD.83.093014;%%
  
  %\cite{Arguelles:2016uwb}
\bibitem{Arguelles:2016uwb} 
  C.~A.~Arg\:uelles, V.~Brdar and J.~Kopp,
  ``Production of keV Sterile Neutrinos in Supernovae: New Constraints and Gamma Ray Observables,''
  Phys.\ Rev.\ D {\bf 99}, no. 4, 043012 (2019)
%  doi:10.1103/PhysRevD.99.043012
  [arXiv:1605.00654 [hep-ph]].
  %%CITATION = doi:10.1103/PhysRevD.99.043012;%%
  
  %\cite{Suliga:2019bsq}
\bibitem{Suliga:2019bsq}
  A.~M.~Suliga, I.~Tamborra and M.~R.~Wu,
  ``Tau lepton asymmetry by sterile neutrino emission -- Moving beyond one-zone supernova models,''
  arXiv:1908.11382 [astro-ph.HE].
  %%CITATION = ARXIV:1908.11382;%%
  
  %\cite{Syvolap:2019dat}
\bibitem{Syvolap:2019dat} 
  V.~Syvolap, O.~Ruchayskiy and A.~Boyarsky,
  ``Resonance production of keV sterile neutrinos in core-collapse supernovae and lepton number diffusion,''
  arXiv:1909.06320 [hep-ph].
  %%CITATION = ARXIV:1909.06320;%%
 
 %\cite{Choubey:2006aq}
\bibitem{Choubey:2006aq} 
  S.~Choubey, N.~P.~Harries and G.~G.~Ross,
  ``Probing neutrino oscillations from supernovae shock waves via the IceCube detector,''
  Phys.\ Rev.\ D {\bf 74}, 053010 (2006)
%  doi:10.1103/PhysRevD.74.053010
  [hep-ph/0605255].

 %\cite{Choubey:2007ga}
\bibitem{Choubey:2007ga} 
  S.~Choubey, N.~P.~Harries and G.~G.~Ross,
  ``Turbulent supernova shock waves and the sterile neutrino signature in megaton water detectors,''
  Phys.\ Rev.\ D {\bf 76}, 073013 (2007)
%  doi:10.1103/PhysRevD.76.073013
  [hep-ph/0703092 [HEP-PH]].
   
  %\cite{Tamborra:2011is}
\bibitem{Tamborra:2011is} 
  I.~Tamborra, G.~G.~Raffelt, L.~Hudepohl and H.~T.~Janka,
  ``Impact of eV-mass sterile neutrinos on neutrino-driven supernova outflows,''
  JCAP {\bf 1201}, 013 (2012)
%  doi:10.1088/1475-7516/2012/01/013
  [arXiv:1110.2104 [astro-ph.SR]].
  %%CITATION = doi:10.1088/1475-7516/2012/01/013;%%
  

  
  \bibitem{Esmaili:2014gya} 
  A.~Esmaili, O.~L.~G.~Peres and P.~D.~Serpico,
  ``Impact of sterile neutrinos on the early time flux from a galactic supernova,''
  Phys.\ Rev.\ D {\bf 90}, no. 3, 033013 (2014)
%  doi:10.1103/PhysRevD.90.033013
  [arXiv:1402.1453 [hep-ph]].
  %%CITATION = doi:10.1103/PhysRevD.90.033013;%%  
  
  %\cite{Dolgov:2000pj}
\bibitem{Dolgov:2000pj} 
  A.~D.~Dolgov, S.~H.~Hansen, G.~Raffelt and D.~V.~Semikoz,
  ``Cosmological and astrophysical bounds on a heavy sterile neutrino and the KARMEN anomaly,''
  Nucl.\ Phys.\ B {\bf 580}, 331 (2000)
%  doi:10.1016/S0550-3213(00)00203-0
  [hep-ph/0002223].
  %%CITATION = doi:10.1016/S0550-3213(00)00203-0;%%
  
  %\cite{Dolgov:2000jw}
\bibitem{Dolgov:2000jw} 
  A.~D.~Dolgov, S.~H.~Hansen, G.~Raffelt and D.~V.~Semikoz,
  ``Heavy sterile neutrinos: Bounds from big bang nucleosynthesis and SN1987A,''
  Nucl.\ Phys.\ B {\bf 590}, 562 (2000)
  % doi:10.1016/S0550-3213(00)00566-6
  [hep-ph/0008138].
  %%CITATION = doi:10.1016/S0550-3213(00)00566-6;%%

\bibitem{Appelquist:2002me} 
  T.~Appelquist and R.~Shrock,
  ``Neutrino masses in theories with dynamical electroweak symmetry breaking,''
  Phys.\ Lett.\ B {\bf 548}, 204 (2002)
 % doi:10.1016/S0370-2693(02)02854-X
  [hep-ph/0204141].
  
  %\cite{Asaka:2005an}
\bibitem{Asaka:2005an} 
  T.~Asaka, S.~Blanchet and M.~Shaposhnikov,
  ``The nuMSM, dark matter and neutrino masses,''
  Phys.\ Lett.\ B {\bf 631}, 151 (2005)
 % doi:10.1016/j.physletb.2005.09.070
  [hep-ph/0503065].
  %%CITATION = doi:10.1016/j.physletb.2005.09.070;%%
  
  %\cite{Asaka:2005pn}
\bibitem{Asaka:2005pn} 
  T.~Asaka and M.~Shaposhnikov,
  ``The nuMSM, dark matter and baryon asymmetry of the universe,''
  Phys.\ Lett.\ B {\bf 620}, 17 (2005)
 % doi:10.1016/j.physletb.2005.06.020
  [hep-ph/0505013].
  %%CITATION = doi:10.1016/j.physletb.2005.06.020;%%
  
  %\cite{Alekhin:2015byh}
\bibitem{Alekhin:2015byh} 
  S.~Alekhin {\it et al.},
  ``A facility to Search for Hidden Particles at the CERN SPS: the SHiP physics case,''
  Rept.\ Prog.\ Phys.\  {\bf 79}, no. 12, 124201 (2016)
%  doi:10.1088/0034-4885/79/12/124201
  [arXiv:1504.04855 [hep-ph]].
  
  %\cite{Chun:2019nwi}
\bibitem{Chun:2019nwi} 
  E.~J.~Chun, A.~Das, S.~Mandal, M.~Mitra and N.~Sinha,
  ``Sensitivity of Lepton Number Violating Meson Decays in Different Experiments,''
  arXiv:1908.09562 [hep-ph].
  %%CITATION = ARXIV:1908.09562;%%
  
  
  %\cite{Fuller:2009zz}
\bibitem{Fuller:2009zz} 
  G.~M.~Fuller, A.~Kusenko and K.~Petraki,
  ``Heavy sterile neutrinos and supernova explosions,''
  Phys.\ Lett.\ B {\bf 670}, 281 (2009)
%  doi:10.1016/j.physletb.2008.11.016
  [arXiv:0806.4273 [astro-ph]].
  %%CITATION = doi:10.1016/j.physletb.2008.11.016;%%
  
% \cite{Rembiasz:2018lok}
%\cite{Rembiasz:2018lok}
\bibitem{Rembiasz:2018lok} 
  T.~Rembiasz, M.~Obergaulinger, M.~Masip, M.~\'A.~P\'erez-Garc\'ia, M.~\'A.~Aloy and C.~Albertus,
  %``Heavy sterile neutrinos in stellar core-collapse,''
  Phys.\ Rev.\ D {\bf 98}, no. 10, 103010 (2018)
  %doi:10.1103/PhysRevD.98.103010
  [arXiv:1806.03300 [astro-ph.HE]].
  %%CITATION = doi:10.1103/PhysRevD.98.103010;%%
  


  %\cite{Mezzacappa:1993gn}
\bibitem{Mezzacappa:1993gn} 
  A.~Mezzacappa and S.~W.~Bruenn,
  ``A numerical method for solving the neutrino Boltzmann equation coupled to spherically symmetric stellar core collapse,''
  Astrophys.\ J.\  {\bf 405}, 669 (1993).
%  doi:10.1086/172395
  %%CITATION = doi:10.1086/172395;%%
  
  %\cite{Liebendoerfer:2002xn}
\bibitem{Liebendoerfer:2002xn} 
  M.~Liebendoerfer, O.~E.~B.~Messer, A.~Mezzacappa, S.~W.~Bruenn, C.~Y.~Cardall and F.~K.~Thielemann,
  ``A Finite difference representation of neutrino radiation hydrodynamics for spherically symmetric general relativistic supernova simulations,''
  Astrophys.\ J.\ Suppl.\  {\bf 150}, 263 (2004)
%  doi:10.1086/380191
  [astro-ph/0207036].
  %%CITATION = doi:10.1086/380191;%%
  
    %\cite{Fischer:2016}
\bibitem{Fischer:2016} 
  T.~Fischer,
  ``The role of medium modifications for neutrino-pair processes from nucleon-nucleon bremsstrahlung - Impact on the protoneutron star deleptonization,''
  Astron.\ Astrophys.\  {\bf 593}, A103 (2016)
 % doi:10.1051/0004-6361/201628991
  [arXiv:1608.05004 [astro-ph.HE]].
  
  %\cite{Raffelt:1987yt}
\bibitem{Raffelt:1987yt} 
  G.~Raffelt and D.~Seckel,
  ``Bounds on Exotic Particle Interactions from SN 1987a,''
  Phys.\ Rev.\ Lett.\  {\bf 60}, 1793 (1988).
%  doi:10.1103/PhysRevLett.60.1793
  %%CITATION = doi:10.1103/PhysRevLett.60.1793;%%
  
    %\cite{Raffelt:2003en}
\bibitem{Raffelt:2003en} 
  G.~G.~Raffelt, M.~T.~Keil, R.~Buras, H.~T.~Janka and M.~Rampp,
  ``Supernova neutrinos: Flavor-dependent fluxes and spectra,''
  astro-ph/0303226.
  %%CITATION = ASTRO-PH/0303226;%%
  
  %\cite{Tamborra:2012ac}
\bibitem{Tamborra:2012ac} 
  I.~Tamborra, B.~Muller, L.~Hudepohl, H.~T.~Janka and G.~Raffelt,
  ``High-resolution supernova neutrino spectra represented by a simple fit,''
  Phys.\ Rev.\ D {\bf 86}, 125031 (2012)
%  doi:10.1103/PhysRevD.86.125031
  [arXiv:1211.3920 [astro-ph.SR]].
  %%CITATION = doi:10.1103/PhysRevD.86.125031;%%
  
  %\cite{Fischer:2009af}
\bibitem{Fischer:2009af} 
  T.~Fischer, S.~C.~Whitehouse, A.~Mezzacappa, F.-K.~Thielemann and M.~Liebendorfer,
  ``Protoneutron star evolution and the neutrino driven wind in general relativistic neutrino radiation hydrodynamics simulations,''
  Astron.\ Astrophys.\  {\bf 517}, A80 (2010)
  %doi:10.1051/0004-6361/200913106
  [arXiv:0908.1871 [astro-ph.HE]].
  %%CITATION = doi:10.1051/0004-6361/200913106;%%
  
%\cite{Raffelt:1992bs}
\bibitem{Raffelt:1992bs} 
  G.~Raffelt and G.~Sigl,
  ``Neutrino flavor conversion in a supernova core,''
  Astropart.\ Phys.\  {\bf 1}, 165 (1993)
%  doi:10.1016/0927-6505(93)90020-E
  [astro-ph/9209005].
  %%CITATION = doi:10.1016/0927-6505(93)90020-E;%% 
  
  %\cite{Hannestad:1995rs}
\bibitem{Hannestad:1995rs} 
  S.~Hannestad and J.~Madsen,
  ``Neutrino decoupling in the early universe,''
  Phys.\ Rev.\ D {\bf 52}, 1764 (1995)
%  doi:10.1103/PhysRevD.52.1764
  [astro-ph/9506015].
  %%CITATION = doi:10.1103/PhysRevD.52.1764;%%
  
  %\cite{Tamborra:2017ubu}
\bibitem{Tamborra:2017ubu} 
  I.~Tamborra, L.~Huedepohl, G.~Raffelt and H.~T.~Janka,
  ``Flavor-dependent neutrino angular distribution in core-collapse supernovae,''
  Astrophys.\ J.\  {\bf 839}, 132 (2017)
%  doi:10.3847/1538-4357/aa6a18
  [arXiv:1702.00060 [astro-ph.HE]].
  %%CITATION = doi:10.3847/1538-4357/aa6a18;%%
  
    %\cite{Sung:2019xie}
\bibitem{Sung:2019xie} 
  A.~Sung, H.~Tu and M.~R.~Wu,
  ``New constraint from supernova explosions on light particles beyond the Standard Model,''
  Phys.\ Rev.\ D {\bf 99}, no. 12, 121305 (2019)
%  doi:10.1103/PhysRevD.99.121305
  [arXiv:1903.07923 [hep-ph]].
  %%CITATION = doi:10.1103/PhysRevD.99.121305;%%
  
  %\cite{Dreiner:2003wh}
\bibitem{Dreiner:2003wh} 
  H.~K.~Dreiner, C.~Hanhart, U.~Langenfeld and D.~R.~Phillips,
  ``Supernovae and light neutralinos: SN1987A bounds on supersymmetry revisited,''
  Phys.\ Rev.\ D {\bf 68}, 055004 (2003)
%  doi:10.1103/PhysRevD.68.055004
  [hep-ph/0304289].
  %%CITATION = doi:10.1103/PhysRevD.68.055004;%%
  
  %\cite{Gelmini:2019deq}
\bibitem{Gelmini:2019deq} 
  G.~B.~Gelmini, A.~Kusenko and V.~Takhistov,
  ``Hints of Sterile Neutrinos in Recent Measurements of the Hubble Parameter,''
  arXiv:1906.10136 [astro-ph.CO].
  %%CITATION = ARXIV:1906.10136;%%
  
  %\cite{Atre:2009rg}
\bibitem{Atre:2009rg} 
  A.~Atre, T.~Han, S.~Pascoli and B.~Zhang,
  ``The Search for Heavy Majorana Neutrinos,''
  JHEP {\bf 0905}, 030 (2009)
%  doi:10.1088/1126-6708/2009/05/030
  [arXiv:0901.3589 [hep-ph]].
  %%CITATION = doi:10.1088/1126-6708/2009/05/030;%%
  
  %\cite{Bondarenko:2018ptm}
\bibitem{Bondarenko:2018ptm} 
  K.~Bondarenko, A.~Boyarsky, D.~Gorbunov and O.~Ruchayskiy,
  ``Phenomenology of GeV-scale Heavy Neutral Leptons,''
  JHEP {\bf 1811}, 032 (2018)
%  doi:10.1007/JHEP11(2018)032
  [arXiv:1805.08567 [hep-ph]].
  %%CITATION = doi:10.1007/JHEP11(2018)032;%%
  
  %\cite{Ballett:2019bgd}
\bibitem{Ballett:2019bgd} 
  P.~Ballett, T.~Boschi and S.~Pascoli,
  ``Heavy Neutral Leptons from low-scale seesaws at the DUNE Near Detector,''
  arXiv:1905.00284 [hep-ph].
  %%CITATION = ARXIV:1905.00284;%%
  
  %\cite{Bryman:2019bjg}
\bibitem{Bryman:2019bjg} 
  D.~A.~Bryman and R.~Shrock,
  ``Constraints on Sterile Neutrinos in the MeV to GeV Mass Range,''
  Phys.\ Rev.\ D {\bf 100}, 073011 (2019)
 % doi:10.1103/PhysRevD.100.073011
  [arXiv:1909.11198 [hep-ph]].
  
  %\cite{Bryman:2019ssi}
\bibitem{Bryman:2019ssi} 
  D.~A.~Bryman and R.~Shrock,
  ``Improved Constraints on Sterile Neutrinos in the MeV to GeV Mass Range,''
  Phys.\ Rev.\ D {\bf 100}, no. 5, 053006 (2019)
%  doi:10.1103/PhysRevD.100.053006
  [arXiv:1904.06787 [hep-ph]].
  %%CITATION = doi:10.1103/PhysRevD.100.053006;%%
  
  
  
  %\cite{Oberauer:1993yr}
\bibitem{Oberauer:1993yr} 
  L.~Oberauer, C.~Hagner, G.~Raffelt and E.~Rieger,
  ``Supernova bounds on neutrino radiative decays,''
  Astropart.\ Phys.\  {\bf 1}, 377 (1993).
%  doi:10.1016/0927-6505(93)90004-W
  %%CITATION = doi:10.1016/0927-6505(93)90004-W;%%
  
    %\cite{Abe:2018uyc}
\bibitem{Abe:2018uyc} 
  K.~Abe {\it et al.} [Hyper-Kamiokande Collaboration],
  ``Hyper-Kamiokande Design Report,''
  arXiv:1805.04163 [physics.ins-det].
  %%CITATION = ARXIV:1805.04163;%%
  
  %\cite{Capozzi:2018ubv}
\bibitem{Capozzi:2018ubv}
  F.~Capozzi, E.~Lisi, A.~Marrone and A.~Palazzo,
  ``Current unknowns in the three neutrino framework,''
  Prog.\ Part.\ Nucl.\ Phys.\  {\bf 102} (2018) 48
%  doi:10.1016/j.ppnp.2018.05.005
  [arXiv:1804.09678 [hep-ph]].
  %%CITATION = doi:10.1016/j.ppnp.2018.05.005;%%
  
  %\cite{Wolfenstein:1977ue}
\bibitem{Wolfenstein:1977ue} 
  L.~Wolfenstein,
  ``Neutrino Oscillations in Matter,''
  Phys.\ Rev.\ D {\bf 17}, 2369 (1978).
%  doi:10.1103/PhysRevD.17.2369
  %%CITATION = doi:10.1103/PhysRevD.17.2369;%%
  
  %\cite{Mikheev:1986gs}
\bibitem{Mikheev:1986gs}
  S.~P.~Mikheyev and A.~Y.~Smirnov,
  ``Resonance Amplification of Oscillations in Matter and Spectroscopy of Solar Neutrinos,''
  Sov.\ J.\ Nucl.\ Phys.\  {\bf 42} (1985) 913
   [Yad.\ Fiz.\  {\bf 42} (1985) 1441].
  %%CITATION = SJNCA,42,913;%%
  
  %\cite{Dighe:1999bi}
\bibitem{Dighe:1999bi} 
  A.~S.~Dighe and A.~Y.~Smirnov,
  ``Identifying the neutrino mass spectrum from the neutrino burst from a supernova,''
  Phys.\ Rev.\ D {\bf 62}, 033007 (2000)
%  doi:10.1103/PhysRevD.62.033007
  [hep-ph/9907423].
  %%CITATION = doi:10.1103/PhysRevD.62.033007;%%
  
  %\cite{Airen:2018nvp}
\bibitem{Airen:2018nvp} 
  S.~Airen, F.~Capozzi, S.~Chakraborty, B.~Dasgupta, G.~Raffelt and T.~Stirner,
  ``Normal-mode Analysis for Collective Neutrino Oscillations,''
  JCAP {\bf 1812}, 019 (2018)
%  doi:10.1088/1475-7516/2018/12/019
  [arXiv:1809.09137 [hep-ph]].
  %%CITATION = doi:10.1088/1475-7516/2018/12/019;%%
  
  
  \bibitem{Mirizzi:2006xx} 
  A.~Mirizzi, G.~G.~Raffelt and P.~D.~Serpico,
  ``Earth matter effects in supernova neutrinos: Optimal detector locations,''
 JCAP {\bf 0605}, 012 (2006)
 % doi:10.1088/1475-7516/2006/05/012
  [astro-ph/0604300].
  
%\cite{Strumia:2003zx}
\bibitem{Strumia:2003zx} 
  A.~Strumia and F.~Vissani,
  ``Precise quasielastic neutrino/nucleon cross-section,''
  Phys.\ Lett.\ B {\bf 564}, 42 (2003)
%  doi:10.1016/S0370-2693(03)00616-6
  [astro-ph/0302055].
  %%CITATION = doi:10.1016/S0370-2693(03)00616-6;%%
  

  
  %\cite{Acciarri:2015uup}
\bibitem{Acciarri:2015uup} 
  R.~Acciarri {\it et al.} [DUNE Collaboration],
  ``Long-Baseline Neutrino Facility (LBNF) and Deep Underground Neutrino Experiment (DUNE) : Conceptual Design Report, Volume 2: The Physics Program for DUNE at LBNF,''
  arXiv:1512.06148 [physics.ins-det].
  %%CITATION = ARXIV:1512.06148;%%
  
  %\cite{An:2015jdp}
\bibitem{An:2015jdp} 
  F.~An {\it et al.} [JUNO Collaboration],
  ``Neutrino Physics with JUNO,''
  J.\ Phys.\ G {\bf 43}, no. 3, 030401 (2016)
%  doi:10.1088/0954-3899/43/3/030401
  [arXiv:1507.05613 [physics.ins-det]].
  %%CITATION = doi:10.1088/0954-3899/43/3/030401;%%
  
  
\bibitem{Chun:2019nwi} 
  E.~J.~Chun, A.~Das, S.~Mandal, M.~Mitra and N.~Sinha,
  ``Sensitivity of Lepton Number Violating Meson Decays in Different Experiments,''
  arXiv:1908.09562 [hep-ph].
  
\bibitem{Gorbunov:2007ak} 
  D.~Gorbunov and M.~Shaposhnikov,
``How to find neutral leptons of the $\nu$MSM?,''
  JHEP {\bf 0710}, 015 (2007)
  Erratum: [JHEP {\bf 1311}, 101 (2013)]
%  doi:10.1007/JHEP11(2013)101, 10.1088/1126-6708/2007/10/015
  [arXiv:0705.1729 [hep-ph]].
  

%\cite{Shrock:1982sc}
\bibitem{Shrock:1982sc} 
  R.~E.~Shrock,
  ``Electromagnetic Properties and Decays of Dirac and Majorana Neutrinos in a General Class of Gauge Theories,''
  Nucl.\ Phys.\ B {\bf 206}, 359 (1982).
%  doi:10.1016/0550-3213(82)90273-5
  %%CITATION = doi:10.1016/0550-3213(82)90273-5;%%
  %202 citations counted in INSPIRE as of 24 Oct 2019  


  
\end{thebibliography}
\end{document}